# Populist Constitutional Backsliding and Judicial Independence: Evidence from Türkiye[1]

Nuno Garoupa      Rok Spruk

## Abstract

*The synthetic control method has emerged as a widely utilized empirical tool for estimating the causal effects of public policies, natural disasters, and other interventions on various economic, social, institutional, and political outcomes. In this study, we demonstrate the potential application of this method in empirical comparative law by estimating the impact of the 2010 constitutional referendum in Türkiye on the trajectory of judicial independence. By comparing Türkiye with a salient Mediterranean donor pool of countries that did not experience similar interventions during the period from 1987 to 2021, we provide evidence of a severe breakdown and erosion of judicial independence. This deterioration appears to be a direct response to the populist constitutional backsliding initiated by the government's assault on the judiciary, which was carried out under the guise of judicial modernization in 2010, before the additional constitutional reforms in 2017.*

**Keywords**: synthetic control method, empirical comparative law, Türkiye, constitutional backsliding, populism, judicial independence

---

[1] Garoupa: Professor of Law, Antonin Scalia Law School, George Mason University, 3301 Fairfax Dr, Arlington, VA 22201, Email: ngaroup@gmu.edu. Spruk (corresponding author): Associate Professor of Economics, School of Economics and Business, University of Ljubljana, Kardeljeva ploscad 17, SI-1000 Ljubljana, Email: rok.spruk@ef.uni-lj.si We are grateful to the participants of ALEA 2023 (Boston), AEDE 2023 (Madrid), and EALE 2023 (Berlin) and Litigation Workshop 2023 (Catania) participants were helpful to the authors. We are grateful to Madeline Conn and Savanah K. Patterson for research assistance. The usual disclaimers apply. The authors jointly declare no conflict of interest. The authors jointly declare no conflict of interest.

# 1  Introduction

Empirical comparative law has slowly taken attention in the field (Spamann 2015, Jacobi 2023, Hertogh 2024). However, compared to the other branches of empirical legal studies, the use of the synthetic control method in comparative law in evaluating both the short-term and long-term effects of small-scale and large-scale legal changes has been scarce and relatively untouched by the empirical credibility revolution that has become a norm elsewhere in social sciences (Angrist and Pischke 2010). To bridge the gap between the wide-standing void in the literature between the theoretical debate and the empirical rigorousness in comparative law, in this article we highlight the importance and discuss the wide potential the application of the synthetic control method into the field of empirical comparative law. We partially fill the void in the literature and demonstrate both the use and application of the method in comparative law by exploiting the populist constitutional reforms unleashed by Türkiye's president Recep Tayip Erdoğan in 2010 after the constitutional referendum.

Although these reforms were ostensibly aimed at aligning the Turkish judiciary with European standards of governance, impartiality, and efficacy, they instead triggered a wave of populist constitutional backsliding, which has been linked to the near-complete erosion of judicial independence. Using a donor pool of Mediterranean states from 1987 to the present, we construct counterfactual scenarios to estimate judicial independence outcomes in the absence of these populist reforms. By comparing Türkiye with its synthetic control group prior to the onset of these reforms, we quantify the extent of judicial erosion and assess its statistical significance. In doing so, we not only demonstrate the tools necessary to evaluate the impact and uniqueness of Türkiye's constitutional backsliding but also situate our findings within recent developments in counterfactual analysis (Abadie and L'Hour 2021, Ben-Michael et al. 2021).

Our findings empirically confirm the widespread and enduring erosion of judicial independence resulting from the ruling party's assault on the judiciary. Compared to its synthetic counterfactual, Türkiye's trajectory reveals a rapid and irreversible decline in judicial independence, with no evidence of merely temporary effects. The estimated negative gaps are statistically significant and robust to a range of internal validity and falsification tests. Moreover, our results suggest that the 2017 constitutional referendum did not serve as a major turning point in this decline—contrary to expectations, the 2010 judicial reforms appear to be the structural break responsible for the erosion, while the 2017 referendum had no perceivable statistically significant effect.

The remainder of the paper is organized as follows. Section 2 discuss the state of the existing literature. Section 3 outlines and presents the identification strategy in synthetic control method.

Section 4 discusses the data and presents both the main results and robustness checks. Section 5 concludes.

## 2  Prior Literature

### 2.1. Synthetic Control Method

The synthetic control method has gained significant momentum in the scholarly literature and has been applied in several different settings across a wide range of applications in economics and political science.[2] As explained by Abadie (2021) and Gilchrist et. al. (2023), the method has been designed for comparative case studies in small and moderately sized samples as a transparent, data-driven way of constructing a synthetic control group for use in comparing the outcome of interest in an affected unit with the outcome of that control group without the direct exposure of that control group to the intervention of interest (Ben Michael et. al. 2021). Data-driven processes of constructing the synthetic control group are based on selecting weights of past outcomes and auxiliary covariates through the diagonal matrix comparison that determines which comparison units' outcome process characteristics are within the convex hull of the affected unit to best reproduce the outcome trajectory for the treated unit in the hypothetical absence of the intervention (Abadie and Cattaneo 2018, Cattaneo et. al. 2021).

### 2.2. Judicial Independence

There is a vast theoretical literature about judicial independence.[3] From an empirical perspective, comparisons across countries have generated different metric approaches (Aydin 2013, Ríos-Figueroa and Staton 2014, Linzer and Staton 2015, Šipulova et. al. 2022) and a long debate about the distinction between *de jure* and *de facto* judicial independence (Hayo and Voigt 2007, 2014, 2016, 2019). The available evidence suggests that populism in power has resulted in severely undermining judicial independence, mostly in three countries: Hungary, Türkiye, and Poland (Gora and de Wilde 2022). As suggested by the leading scholars in the field of comparative law and politics (Varol 2016, Ginsburg et. al. 2018, Scheppele 2018), the erosion of judicial independence has been intrinsic to democratic backsliding in the last decades. In power, populist governments attempt to undermine judicial institutions and civil liberties as part of their effort to promote an alternative model to liberal democracy. They operate through law and legal reform. Different authors have used terms such as autocratic legalism, constitutional retrogression, or abusive constitutionalism to explain these political processes (de Sa e Silva 2022). There is a strong association between the rise of populism and the decline of the rule of law (Lacey 2019, Gora and de Wilde 2022). However, from a methodological perspective, these empirical findings rely on short-term correlations or associations between populism takeover and decline in judicial independence indicators. More importantly, the causal mechanism is highly debatable and certainly not possible to be established or tested by descriptive statistics and simple linear correlations.

---

[2] See the long list of applications summarized by Abadie 2021, Gilchrist et al 2023 and, specifically, Abadie et al. 2015 in the field of comparative politics.
[3] For example, see generally Landes and Posner 1975, Larkins 1996, Howard and Carey 2004, Helmke and Rosenbluth 2009, Epperly 2019. There is a voluminous qualitative (Ginsburg 2003) and quantitative (Epperly and Sievert 2019, Dijk 2021, Hertogh 2024) literature about judicial independence.

An important paper by Aydin-Cakir (2024) used synthetic control method to explore the impact of different forms of judicial reform (formal and informal mechanisms) by populist governments on judicial independence. More specifically, the author compared judicial independence levels for Hungary, Poland, and their synthetic controls. She suggested that, in the Polish case, the court-curbing attempts had greater negative impact on judicial independence than the court-curbing attempts used in Hungary.

### 2.3. The Case of Türkiye

There is, by now, an enormous academic literature discussing how populist governments have presided over a decline in judicial independence in countries such as Türkiye in the last two decades (Aytaç and Elçi 2018, Kadıoğlu 2021, Kurban 2024). In Türkiye, the approval of the constitutional amendments of September 2010 represented the turning point in the tense relationship between the courts and executive power. Initially, these reforms were presented as bringing Türkiye in line with European practices (Özbudun 2011). However, they created the critical opportunity to pack the Turkish Constitutional Court with appointees of Erdoğan's AKP, the Justice and Development Party (Yeginsu 2010, Cook 2016, Varol et. al. 2017, Aytaç and Elçi 2019). The reform of the Turkish Constitutional Court (AYM) and the reconfiguration of the High Council of Judges and Prosecutors (HSYK), in charge of appointments, promotions, transfers, and suspensions of all judges and public prosecutors, played an important role in the campaign for the constitutional referendum (Kalaycıoğlu, 2012). The amendments increased the size of the AYM from eleven to seventeen members, thus allowing the governing AKP to appoint new justices with the goal of taming a predominantly secularist Constitutional Court (Varol et. al. 2017). Furthermore, these amendments also enlarged the composition of the HSYK, including members elected among their peers, thus paving the way for the government to pack the HSYK with sympathetic members, following internal elections where intimidation prevailed (Kalaycıoğlu 2012).

In late 2013, another next major step undercutting the judiciary's independence took place. The Turkish prosecutors indicted businesspeople and AKP politicians for corruption. The government responded by removing and reassigning hundreds of judges and prosecutors from their posts, as well as by further reorganizing the HSYK in such a way as to effectively subjugate it to the Ministry of Justice (Müftüler-Baç 2019, Tahiroglu 2020, Soyaltin-Colella 2022). Following the failed coup attempt of 2016, the state of emergency was used by the government to purge more than one-quarter of the judges and prosecutors, two constitutional court justices, and hundreds of staff members in the high courts (Felter and Aydin 2018, Tahiroglu 2020). By the end of this process (the 2017 Turkish constitutional reform referendum), according to scholars, the Turkish judiciary was largely domesticated and heavily controlled by the executive branch of government (Tahiroglu 2020).[4]

## 3    Identification strategy

### 3.1    Basic setup

The primary objective of our identification strategy is to consistently assess the impact of the populist constitutional reforms introduced by the Erdoğan administration in 2010 on judicial quality and independence. Central to our approach is the estimation of the missing counterfactual scenario linked to the 2010 constitutional referendum and the resulting decline in judicial independence. To this end, our empirical strategy focuses on tracing the trajectory of judicial independence over time, in the hypothetical absence of these populist reforms. In contrast to the more widely used difference-in-differences approach, which primarily estimates the average treatment effect, our key variable of interest is the dynamic evolution of judicial independence up to the present day. The reliability of the difference-in-differences method depends on the assumption of parallel trends between treated and control units. If this assumption is violated, the estimated average treatment effect may be compromised, raising concerns about the validity and interpretability of the results. Unlike difference-in-differences, the synthetic control method does not necessarily require parallel trends. Instead, synthetic control method constructs control groups through a two-stage process involving both training and validation periods, ensuring that outcome and covariate-level similarities are properly accounted for. This makes synthetic control method a more robust tool for estimating long-term changes in judicial independence in response to Türkiye's constitutional reforms. Therefore, to estimate the counterfactual scenario we follow the established and well-known empirical strategy set out by Abadie and Gardeazabal (2003), Abadie et. al. (2010, 2015) and further improved by Billmeier and Nannicini (2013), Cavallo et.

---

[4] These amendments proposed the abolition of the parliamentary system of government and its replacement by a presidential system, the increase in the number of seats in the parliament from 550 to 600, and substantially expanding the presidential overreach over appointments to the HSYK. It should be noted that the referendum was held under the state of emergency declared following the failed military coup d'état by a faction within the Turkish Armed Forces in July 2016. The referendum took place in April 2017 and was narrowly approved after 51 percent of the votes were casted in favor of the proposed amendments.

al. (2013), Xu (2017), Cerulli (2019), Ben-Michael et. al. (2021), , Kaul et. al. (2022), and Gilchrist et. al. (2023) amongst several others.

We observe a series of judicial independence trajectories of $(J \times 1) \in N$ countries in the time period $t = 1, 2, \ldots T$. The constitutional reform takes place at time $T_0 \in (1, T)$ such that $T_0 < T$ and affects only Türkiye which is denoted as $J - 1$-th country and lasts from period $T_0 + 1$ until $T$ without interruption. In addition, let $Q_{j,t}^N$ be the potential trajectory of judicial independence as a realization of the scenario in the eventual absence of the constitutional reforms. Analogically, let $Q_{j,t}^I$ represent the observed trajectory of judicial independence in the presence of constitutional reforms. Borrowing the terminology from treatment effects literature, our key parameter of interest is the treatment effect of 2010 constitutional reforms. Without the loss of generality, the underlying treatment effect can be written as:

$$\gamma_{j,t} = Q_{j,t}^I - Q_{j,t}^N \tag{1}$$

where $Q_{j,t}^I$ is the observed realization of judicial independence indicator in j-th country at time t, and $Q_{j,t}^N$ is the missing counterfactual scenario of judicial independence without the populist constitutional reforms. Furthermore, the constitutional reforms may be described as a binary variable $D$ that can take the value of 1 for the period $t \geq T_0$ and 0 for the period $t < T_0$. Given that Türkiye is the singular treated country in our setup with an innate implication that it is indexed as $j = 1$, the observed outcome realization in the presence of populist reforms is given by:

$$Q_{j,t} = Q_{j,t}^N + \hat{\gamma}_{j,t} \cdot D_{j,t} \tag{2}$$

$$D_{j,t} = \begin{cases} 1 & \text{if } j = 1 \land t \geq T_0 \\ 0 & \text{otherwise} \end{cases} \tag{3}$$

where we estimate the entire vector of post-treatment effects associated with the constitutional reforms which can be described as the parametric sequence $\{\gamma_{1,T_0+1}, \ldots \gamma_{1,T}\}$. Each element within the vector of post-treatment effects captures the influence of constitutional reforms on the trajectory of judicial independence from the period $T_0$ until $T$. Per se, $Q_{j,t}^I$ is observable across space and time for the period $t > T_0$ whilst $Q_{j,t}^N$ has to be estimated to uncover the projection of the counterfactual scenario in the hypothetical absence of reforms. Henceforth, let $Q_j = [Q_{j,1} \ldots Q_{j,T_0}]$ denote pre-reform vector of observed judicial independence trajectories for country $j \in \{1, \ldots J + 1\}$, and let $\mathbf{X_j}$ be a $(K \times 1)$ vector of auxiliary covariates of $\mathbf{Q_j}$. These two distinctive vectors allow us to build $\mathbf{Q_0} = [Q_2 \ldots Q_{J+1}]$ and $\mathbf{Y_0} = [Y_2 \ldots Y_{J+1}]$ matrices invertible across $(K \times J)$ dimension that contain the outcome and covariate values in pre-treatment period for the countries not affected by the populist constitutional reforms at time $T_0$ and beyond.

In the lieu of the binary nature of the treatment, the potential outcome scheme for the judicial independence is given by the following simple latent factor model:

$$Q_{j,t}^{D \in \{0,1\}} = \begin{cases} Q_{j,t}^N (\text{without reforms} \mid D=1) = \phi_t + \mathbf{Z}_{j,t}'\theta_t + \lambda_t \mu_j + \varepsilon_{j,t} \\ Q_{j,t}(\text{with reforms} \mid D=0) = \quad\quad Q_{j,t}^N + \hat{\gamma}_{j,t} \cdot D_{j,t} \end{cases} \quad (4)$$

where $\phi_t$ represents unobserved time-varying technology shocks to judicial independent common to all countries, $\mathbf{Z}_{j,t}$ is an $(1 \times r)$ vector of observable auxiliary covariates of judicial independence, $\theta_t$ is an $(r \times 1)$ vector of unknown parameters, $\lambda_t$ is an $(1 \times F)$ vector of unknown common factors, and $\mu_j$ is an $(F \times 1)$ vector of unknown factor loadings. The transitory shocks to judicial independence are given by $\varepsilon_{j,t}$ and are assumed to be identically and independently distributed such that $\varepsilon_{j,t} \sim i.i.d$ and $E\left(\varepsilon_{j,t} \mid D(1,T_0)\right) = 0$. It should be noted that the key parameter of interest is $\lambda_t \mu_j$ which allows us to capture the temporal heterogeneity of the response to the constitutional reforms which does not necessitate parallel trend assumption to hold.

Therefore, our goal is to construct a series of artificial control groups for Türkiye being inasmuch as possible similar to the observed trajectories of judicial independence. More specifically, by reweighing the judicial independence trajectories of the countries using the implicit $\mathbf{Q_0}$ and $\mathbf{Y_0}$ characteristics from the corresponding donor pool in pre-$T_0$ period, the trajectory of $Q_{j,t}^N$ can be estimated for each year in the pre- and post-treatment period denoted as $t \in \{1, \dots T\}$. Under these circumstances, $Q_{j,t}^N$ consists of the reweighted combination of the implicit and explicit attributes of countries from the respective synthetic control group which implies that:

$$\hat{Q}_{j,t}^N = \sum_{j=2}^{J+1} \hat{w}_j Q_{j,t} \quad (5)$$

where $\widehat{\mathbf{W}} = [w_2, \dots w_{J+1}]' \in \mathbb{R}$ describes the vector of weights used to construct the series of synthetic control groups for Türkiye in the hypothetical absence of the populist constitutional reforms. To approximate the trajectory of the counterfactual scenario, $\widehat{\mathbf{W}}$ can be estimated by finding the solution to the following single-nested minimization problem:

$$\widehat{\mathbf{W}}(\widehat{\mathbf{V}}) = \underset{\mathbf{W} \in \mathcal{W}}{\operatorname{argmin}} (\mathbf{X_1} - \mathbf{X_0 W})' \mathbf{V} (\mathbf{X_1} - \mathbf{X_0 W}) \quad (6)$$

where the vector of weights $W = (w_1, \dots w_j)$ is restricted to be non-negative, convex and additive which implies that $w_j \geq 0$ for each $j \in \{2, \dots J+1\}$ and $\sum_{j=2}^{J+1} w_j = 1$.[5] It should be noted that a

---
[5] It should be noted that the convexity requirement is a sufficient but not a necessary condition to estimate the set of weights used to build the counterfactual trajectory. To fill the void in the literature, Ben Michael et. al. (2021) propose an augmented version of synthetic control method in settings when pre-treatment quality of the fit is infeasible. More specifically, they propose a bias correction for inexact matching to de-bias the original synthetic control estimate and apply a ridge regression to model the outcomes. Under this approach, the solution to Eq. (6) is bounded on the

positive weight computed for *j*-th country from the donor pool suggests that the treated country falls within the convex hull of the implicit outcome- and covariate-specific hull of the *j*-th country where higher weight share indicates a greater degree of similarity therein. Two matrices are built to solve the nested optimization problem and derive the weights to approximate the similarity between Türkiye and the countries from the donor pool. First, **V** represents a diagonal positive semi-definite matrix of $(K \times K)$ dimension with the sum of main diagonal elements that equal one. More precisely, **V** denotes the explanatory power and importance of pre-$T_0$ outcomes and auxiliary covariates in explaining and predicting $Q$ as the outcome of interest. The matrix is constructed and obtained in the training stage where the synthetic control algorithm learns the best model specification to explain the outcome of interest. In the validation stage, the vector of weights **W** is formed denoting the set of countries from the donor pool with an additive structure equal to one, which fall within the identified convex hull of Türkiye's judicial independence attributes in the full pre-treatment period. By default, these two matrices are invertible and ensure that the designated synthetic control group mimics Türkiye's trajectory of outcomes inasmuch as possible. The exact choice of **V** matrix has been a subject of rigorous debate where a variety of solutions has been proposed (Gobillon and Magnac 2016, Hahn and Shi 2017, Robbins et. al. 2017, Amjad et. al. 2018, Kaul et. al. 2022, Pang et. al. 2022).

Against the backdrop, our approach in the choice of **V** is similar to Abadie et. al. (2015), Klößner et. al. (2018), Firpo and Possebom (2018) and Ferman et. al. (2020). In particular, **V** is chosen through a two-stage procedure. In the initial training period $\widehat{\mathbf{W}}(\widehat{\mathbf{V}}) = \underset{\mathbf{W} \in \mathcal{W}}{\operatorname{argmin}}(\mathbf{X_1} - \mathbf{X_0W})'\mathbf{V}(\mathbf{X_1} - \mathbf{X_0W})$ is minimized whereupon we adopt Vanderbei (1999) constrained quadratic optimization routine. This routine is based on a simple algorithm using interior point method to solve the quadratic programming problem under the imposed constraints. The implementation of the method takes place vis-á-vis C++ plugin where the standard tuning parameters are imposed such as 5% constraint for the tolerance of violation, the maximum number of iterations is set at 1,000 and the clipping bound for the variables is set to 10. In the validation stage, the choice of $\widehat{\mathbf{W}}(\mathbf{V})$ is cross-validated to minimize the out-of-sample prediction error through the follow optimization problem:

$$\widehat{\mathbf{V}} = \underset{\mathbf{V} \in V}{\operatorname{argmin}}(\mathbf{Q_1} - \mathbf{Q_0W(V)})'(\mathbf{Q_1} - \mathbf{Q_0W(V)}) \tag{7}$$

where **V** is a diagonal positive semi-definite invertible matrix of $(K \times K)$ dimension with $\operatorname{tr}(\mathbf{V}) = 1$. More specifically, the matrix of $\mathbf{Q_1}$ is projected on $\mathbf{X_1}$ by imposing $v_k = \frac{|\beta_k|}{\sum_{k=1}^{K}|\beta_k|}$ which denotes

---

estimation error which also allows for negative weights and the extrapolation outside the convex hull of the treated unit.

the k-the diagonal element of **V** and $\beta_k$ is the k-th coefficient of the linear projection of $\mathbf{Q_1}$ on $\mathbf{X_1}$.

### 3.2   Scope of treatment

The synthetic control method offers significant potential for addressing some of the most pressing questions in empirical comparative law by constructing counterfactual scenarios. To effectively apply this method, the key question is: What constitutes an internally valid synthetic control analysis? To identify the policy treatment of interest, the selection process should meet three critical and mutually inclusive criteria. First, the treatment assigned to the affected country or unit should not be easily predicted by pre-existing conditions such as institutional quality or judicial independence. If these innate conditions consistently predict the timing of the treatment, the conditional independence assumption may be violated, undermining the validity of the analysis. The key to treatment selection lies in its relative uniqueness. If the policy intervention is widespread across multiple countries at the same time, these countries should be excluded from the donor pool to ensure a valid and credible counterfactual scenario. For example, if the synthetic version of Türkiye—constructed in the hypothetical absence of the 2010 constitutional reforms—were composed of countries undergoing similar populist backsliding, such as Hungary, Poland, or Russia, the control group would likely fail to provide a meaningful representation of the counterfactual trajectory of judicial independence. Ensuring that the control group reflects distinct conditions from the treated country is crucial for generating an accurate and insightful counterfactual.

### 3.3   Placebo analysis

#### 3.3.1   In-space placebo analysis

Perhaps the most straightforward question arising from estimating the gap between Türkiye as the country affected by populist constitutional backsliding and its synthetic control group in the follow-up period after the reforms concerns the statistical significance of the estimated gap. Since the synthetic control analysis is a non-parametric technique, statistical significance cannot be conducted using conventional parametric inference and test statistics. Instead, to determine whether the vector of estimated post-reform gaps between Türkiye and its synthetic control group is statistically significant or not, we rely on the treatment permutation procedure that has been proposed in the policy evaluation literature by Abadie et. al. (2010), and has been furthered by Hahn and Shi (2017), Ferman and Pinto (2021), Firpo and Possebom (2018) and Chernozhukov et. al. (2021), among several others.

Under the treatment permutation, the respective policy of interesting is assigned to the full set of countries in the donor pool that never underwent the respective policy change in the period of investigation. Once the policy is assigned to other countries, the synthetic control estimator is iteratively applied to each respective country in the donor pool whilst shifting the treated country into the donor pool. Through a series of iterative runs, the distribution of placebo effects is built which can be compared to the estimated gap of the treated country. In brief quantitative terms, treatment permutation and placebo analysis can be described as follows. For each country $j \in \{2, \ldots J+1\}$ in the period $t \in \{1, \ldots T\}$, a full vector of post-treatment effects designated as $\hat{\gamma}_j = \{\hat{\gamma}_{1,T_0+1}, \ldots \hat{\gamma}_T\}$ is estimated upfront. In the subsequent step, the full distribution of placebo effects' vector is built through the permutation of the treatment-related policy to the unaffected countries, and is compared with the empirical distribution of the full treatment effect for Türkiye as a treated country of interest, denoted as $\hat{\gamma}_1 = \{\hat{\gamma}_{1,T_0+1}, \ldots \hat{\gamma}_{1,T}\}$.

The general rationale of such placebo analysis is both simple and straightforward. If the estimated series of judicial independence gaps in response to the populist constitutional reforms for Türkiye is like the gaps in the placebo distribution, the notion of significant effect of such reforms would be questionable given that the countries in the donor pool follow similar trajectories of outcomes as Türkiye. In such circumstance, it is likely that a common shock roughly experienced by a large group of countries in the donor pool would preclude the inference of significant effect of constitutional reforms on judicial independence. By contrast, if the estimated judicial independence gap for Türkiye is unique, imperceptible elsewhere in the donor pool, and large, then our analysis may well provide some tentative evidence of the significant effect of populist constitutional reforms as such notion then becomes more credible and plausible. The underlying null hypothesis behind the estimated effect of populist constitutional reforms is rejected if the vector of full treatment effects for Türkiye is unusually large and the fraction of countries with similar magnitude of effects and the same sign of the gap is less than in the range between 10 percent and 15 percent.

In comparison with the empirical placebo distribution $|\hat{\gamma}_{j,t}|$, the distribution of $|\hat{\gamma}_{1,t}|$ can be abnormally large in some period after the policy intervention but not in the full period which implies that the exact decision rule behind the rejection of the null hypothesis of zero effect may be difficult. A simple and plausible approach to overcome such innate ambiguity has been advocated by Abadie et. al. (2010) to compare the root mean square prediction error of the treated unit (i.e. Türkiye) with the prediction errors in the placebo simulation before and after the policy change. The rationale behind such comparison is eloquent and straightforward. If the effect of the policy of interest is specific to Türkiye as the affected country and is only meagrely perceivable elsewhere, the ratio of Türkiye's RMSE and the RMSE of the empirical placebo distribution in the period after the intervention should be exceptionally small. The prediction error ratio can be computed as:

$$RMSE_j = \left(\frac{\sum_{t=T_0+1}^{T}(Q_{j,t}-\hat{Q}_{j,t}^N)^2}{T-T_0}\right) \div \left(\frac{\sum_{t=T_0}^{T}(Q_{j,t}-\hat{Q}_{j,t}^N)^2}{T-T_0}\right) \qquad (8)$$

where $RMSE_j$ denotes the ratio of mean square prediction errors. The respective ratio can be used to compute non-parametric p-values proposed by Cavallo et. al. (2013) and formally validated by Galiani and Quistorff (2017) which denotes the proportion of units from the donor pool having the RMSE at least as large as the RMSE of the treated unit:

$$\mathbb{P} = \frac{\sum_{j=1}^{J+1} 1 \cdot [RMSE_j \geq RM\quad_1]}{J+1} \qquad (9)$$

where $1[\cdot]$ is a simple Iversionian dichotomous function, $RMSE_j$ is the root mean square prediction error ratio of j-th unit from the donor pool, and $RMSE_1$ is the counterpart root mean square prediction error of the treated unit. Without the loss of generality, the computed p-value cannot be interpreted in the standard parametric framework through the lens of testing sharp null hypothesis. Instead, it may be interpreted as the proportion of units in the donor pool having the estimated effect of the policy change at least as large as the treated unit. Hence, if the proportion of units is high, the null hypothesis of zero effect cannot be rejected. On the other hand, if the proportion is low and within some specified significance threshold such as 0.15, the null hypothesis can be more easily rejected. Henceforth, if the quasi p-value from treatment permutation is sufficiently low, the notion that the estimated gap for the treated unit was obtained either by fluke, chance or at random would not seem plausible and can be refuted. Furthermore, Firpo and Possebom (2018) provided and discuss sufficient conditions that guarantee the formal validity of the inference alongside the size and power of the permutation test.

### 3.3.2 In-time placebo analysis

An additional caveat behind the estimated judicial independence gap in response to the populist constitutional reforms emanates from the choice of the treatment year. A pioneering example has been set out by Abadie et. al. (2015) in the study of the economic growth effect of German unification in 1990s on the West Germany drawing on the earlier literature on non-experimental program impact evaluation (Heckman and Hotz 1989) One extant possibility predating the choice of the treatment year to estimate the effect of policies concerns the confluence of alternative policies and shocks distinctive from the choice of postulated policy. The general thrust behind these comparisons is relatively simple. Namely, if the estimated erosion of judicial independence in response to the populist constitutional reforms is anticipable by distinctive structural breaks taking place in the preceding years, the credibility of the synthetic control estimates and their internal validity can be jeopardized and brought into question (Abadie 2021).

Leveraged against the baseline results, these concerns can be at least partially addressed by conducting an in-time placebo analysis. Contrary to the in-space placebo analysis where policy is assigned to the units in the donor pool, the timing of the constitutional reforms is assigned to a deliberately wrong year by backdating the intervention into the pre-treatment period. Such falsely assigned policy year may be used to gauge temporal placebo analysis and can be seen as falsification test. The confidence in the internal validity of synthetic control estimator would disappear if the method estimated large and similar effects when backdated to the years in which the policy did not take place. By contrast, if a large effect is found for the populist constitutional reforms but no effect is perceptible when the reform period is artificially assigned and backdated to the pre-treatment period, the confidence that the effect estimated for populist constitutional reforms provides a reasonably accurate prediction of trajectories of outcomes for Türkiye becomes more plausible and less susceptible to the alternating influence of distinctive shocks or policies. Additional checks on the choice of the falsely assigned policy year supported by the battery of structural break test are elaborated in Keseljevic and Spruk (2023).

### 3.4 Sensitivity analysis

#### 3.4.1 Leave-one-out analysis

Another possible concern behind the internal validity of the estimated policy impacts using synthetic control method arises from the detailed composition of synthetic control groups. It may happen that the synthetic control groups appear to be relatively evenly balanced across several units from the donor pool as it is equally likely that certain units exert high leverage and disproportionately influence the overall composition of synthetic control group. Given that the composition of control groups shapes the direction of estimated counterfactual scenario, an overly large and excessive influence of particular units may exert substantial uncertainty behind the magnitude of the policy effect that render itself pervasive. More recently, Abadie (2021) argued that the credibility of the synthetic control method critically depends on the ability to quantitatively track and reproduce the trajectory of the outcome variable for the treated unit. Against this backdrop, Klößner et. al. (2018) derived a heuristic rule of thumb to address the ambiguity behind the number of donor units that synthesize the treated unit of interest through the leave-one-out analysis where the robustness of the estimated gap is assessed against the exclusion of individual units from the donor pool with non-zero weight that best synthesize the outcome trajectory of the treated unit. The evidence indicating little ambiguity behind the baseline effect with few instances of deviation through the varying composition of synthetic control groups typically does not at least partially vindicates the ambiguity concerns behind the ability of individual comparison units to taint and critically influence the main effect of the policy.

### 3.4.2 Specification search

One of the most important steps in undertaking the synthetic control analysis of the policy of interest rests on identifying the relevant auxiliary covariates of the outcome of interest. In the ideal circumstances, these covariates can account for the sizeable share of the cross- and within-unit outcome variable in the pre-treatment period. Henceforth, the obvious question to ask is how to properly select the covariates to build a compact and, yet parsimonious synthetic control specification whilst minimizing the risk of omitted variable bias. In providing the background for further discussion, Hahn and Shi (2017) advocated the use of a large number of auxiliary covariates as a way of improving the nested optimization through which cross-unit comparisons are drawn and the optimal weights are selected. Instances of time-varying and time-invariant variables deployed into the synthetic control specification can be found across a broad spectrum of the scholarly literature (Kaul et. al. 2022, Powell 2022). A potentially appealing candidate variable for the selection into the specification is the past outcome dynamics which can be operationalized through past outcome values in specific benchmark years (Abadie et. al. 2010) or using lagged outcome values (Doudchenko and Imbens 2016) although Athey and Imbens (2006) showed that including covariates other than lagged outcome renders their explanatory power almost meaningless. Furthermore, Botosaru and Ferman (2019) and Ferman et. al. (2020) suggested to partially redress the lack of clearly-defined theoretical guidance behind the specification-searching strategies, the use of different sets of lags and covariates along with comprehensive reports should be considered in addition to the discarding of specifications using the average of pre-treatment outcome given its failure to fully exploit the pre-treatment outcome variation and dynamics (Ferman 2021). Leveraging a wide array of specification searches, McClelland and Gault (2017) advocated the choice of a small number of outcome variable lags that most closely track and plausibly reproduce the outcome trend and dynamics in the full pre-treatment period.

To fill the void in the literature on the specification search, Xu (2017) proposed a generalized synthetic control estimator that unifies the standard synthetic control method with a class of linear fixed-effects model allowing for rich cross-effect interaction. More specifically, the generalized synthetic control method imputes the counterfactuals either for single or multiple treated units by extracting a rich information set based on the linear interactive fixed-effects model that jointly accommodates unit-specific intercepts with time-varying coefficients. Henceforth, the ability of the generalized synthetic control estimator to reproduce the counterfactual scenario rests on the incorporation of the unobserved heterogeneity bias and common outcome-related technology shocks that vary over time. This implies that treatment-related policy may exhibit non-zero correlation with the unobserved spatial and temporal heterogeneity with marked improvement in both efficacy and interpretability of the estimates. Furthermore, it also incorporates a cross-validation scheme that can select the number of factors of the interactive-fixed effects model without an arbitrary search which tends to reduce the risk of overfitting provided that the model is correctly specified. The generalization of the synthetic control method to multiple treated units allows the researcher to estimate the interactive fixed-effects model only once and generate counterfactuals in a single run which is particularly suitable and appealing in samples where small number of observations naturally poses a source of strong sensitivity to the idiosyncratic effects. In addition, uncertainty of the estimates can be evaluated through the computation of standard error and the respective confidence intervals which do not per may not necessitate extensive placebo analyses in comparison with the standard synthetic control estimator

### 3.5  Differential trend analysis

A different but complementary aspect of the synthetic control analysis concerns the outcome dynamics before and after the policy intervention of interest. Whilst the main interest behind the analysis lies in whether the change in the outcome can be attributed to the treatment-related policy of interest, the most natural question to ask is whether the outcomes of the treated unit and its synthetic control group follow differential and statistically significant trend in the post-intervention period. In the ideal circumstances, the outcome trend between the treated unit and its synthetic control group should be statistically indistinguishable in the pre-treatment period in terms of slope and intercept. If the effect of the treatment-related policy is internally valid, the outcome trajectories of the treated unit and its control group should exhibit a differential and statistically significance change of the trend in the post-treatment period. In the seminal analysis, Spruk and Kovac (2020) tested for the similarity of outcome trends between the treated unit and its synthetic control group in the analysis of the impact of trans-fats ban in Denmark on obesity and cardiovascular mortality rates. They construct a triple-differences test of the structural break that leverages change in the outcomes against the trend of the change before and after the policy intervention by embedding an exact Fisher test statistic in Chow test framework to determine whether the policy intervention induced the structural break in the trend of the outcome trajectory. They also show that the null hypothesis on the absence of differential trends can be evaluated through a parametric inference procedure based on the two-side test statistics and the corresponding p-value.

Furthermore, 95% confidence intervals and standard errors can be constructed for the outcome change trend slopes before and after the policy intervention. It should be noted that differential trend analysis conveys several advantages. First, it can uncover the presence or absence of differential trend in the outcome change between the treated unit and its control group through the test of sharp null hypothesis. The failure to reject null hypothesis may indicate a relatively weak effect of the policy whereupon the outcome is affected but the trend is left intact. By contrast, the rejection of the null hypotheses may lend further empirical support for the notion that the policy of interest not only affected the outcome of interest but also produced a differential trend in the outcome of the treated unit, testifying to a much stronger and possibly causal effect of the policy. Nonetheless, the ability to empirically assess the differential trend assumption may help alleviate and allay the ambiguities behind the notion whether the policy or intervention of interest appears to have had either a causal or noticeably weaker impact.

### 3.6 Bias-correction analysis

An additional issue in the inference on the effect of the intervention on the outcome of interest is posited by the problem of finding the synthetic control group that best reproduces the attributes and characteristics of the treated unit. Such synthetic control group may not have a unique solution as there may be a multiplicity of different solutions to the nested optimization problem in Eq. (6). In the setup where several units are treated and even more units are untreated, the multiplicity of solution is a pressing challenge. Against this backdrop, Abadie and L'Hour (2021) proposed a more nuanced version of synthetic control estimator where pairwise discrepancies between the attributes of the treated unit and the units that contribute non-zero weights to the synthetic control group are penalized. More specifically, they introduce a penalized parameter trading-off pairwise matching discrepancies with respect to the characteristics of each unit in the synthetic control group against similar discrepancies of the control group as a whole. They show that the penalized estimator is both unique and sparse, and propose a bias-correction strategy to improve and extend the inferential methods to multiple treated units building on the earlier penalization schemes for synthetic control and related methods by Doudchenko and Imbens (2016), Amjad et. al. (2018), Arkhangelsky et. al. (2021), Chernozhukov et. al. (2021). Relatedly, Athey et. al. (2021) proposed an underlying sparse factor structure for the outcomes in the absence of the treatment-related intervention, and propose a variety of matrix completion techniques to estimate the counterfactual scenario. In comparison with Abadie and L'Hour (2021), their estimator does not penalized pairwise discrepancies but the entire complexity of the structure in the latent factor model from Eq. (4). In estimating and evaluating the effects of treatment-related interventions, it is highly recommended that scholars do carefully apply a variety of estimates using different estimators and avoid cherry-picking selected estimator in evaluating the effects of the policies, reforms or related interventions. Additional cautionary limitations and prospective mitigations are discussed by Klößner and Pfeifer (2018), Cattaneo et. al. (2021), Gilchrist et. al. (2023), and Pang et. al. (2022), among several others.

# 4 Application: estimating the effect of populist constitutional reforms on judicial independence in Türkiye, 1987-2021

## 4.1 Data

### 4.1.1 Dependent variables

We collect the data on judicial independence by compiling a variety of indicators using the recently updated Varieties of Democracy dataset (Coppedge et. al. 2022). The dataset collates more than 480 indicators and transforms into five core indices along with other supplementary indices including the independence of judiciary.[6] Six different indicators of judicial independence are considered as the outcomes of interest in our investigation on the annual basis and included in the vector of dependent variables. The variables have been created from large-scale surveys of country-level experts and were transformed onto the ordinal scale using the Bayesian item response theory measurement model. We will refer to these six indicators as judicial independence, although different scholars can quibble with the nature of one or more indicators. For example, Aydin-Cakir (2024) focused exclusively on *high court independence* in her study about Hungary and Poland. Since the empirical results are presented for each indicator individually, one can leave to the reader the decision about which indicators are more or less compelling as appropriate and relevant measures of *de facto* judicial independence.

(a) *High court independence* variable reflects the independence and autonomy of the judiciary in the decision-making process without the interference of fear of the executive branch of government. Thus, the variable can identify the autonomous judicial-decision making and its absence. It can be compared across space and time. By default, judicial decisions can reflect government preferences and wishes as a court can have a high degree autonomy whilst its decisions support the government's position. Judges can also be persuaded about the merits of the government's position. The variable reflects whether the court simply adopts the government adoption with no regard to the sincere view of the government record. The variable can take five distinctive values: (i) zero if the court always adopts the government's position, (ii) one if the high court usually adopts the government's position, (iii) two if the high court adopts the government's position about one half of the time, (iv) three if the high court rarely adopts the government's position, and (v) four if the court never adopts the government's position. Higher values of the variable thereby indicate a somewhat greater and more resilient degree of judicial independence.

---

[6] We are aware of the ongoing debate in political science about the shortcomings of the Varieties of Democracy dataset. Although there are possible criticisms to the variables in terms of measuring perceptions rather than actual changes, the debate is inconclusive at the moment. In the extent that our article is about proposing an empirical methodology, it can be easily applied to an alternative set of indicators.

(b) *Court packing* variable indicates the degree to which the executive branch of government promulgates politically motivated judicial appointments to the high and lower court. The size of the judiciary is at times increased for certain reasons such as the increasing caseload or purely for political reasons when the executive branch aims at influencing the judiciary. The variable can take four distinctive values: (i) three if judgeships were added to the judiciary, but there is no evidence that the increase was politically-motivated and there was simply no increase, (ii) two if judgeships were added to the judiciary and there was a limited politically-motivated increase in the number of judgeships, (iii) one if there was a limited, politically-motivated increase in the number of judgeships on very important courts, and (iv) zero if there was a massive, politically-motivated increase in the number of judgeships across the entire judiciary. Therefore, lower values of the variables on the ordinal scale indicate a greater interference and active involvement of the executive branch of government in the judicial appointments.

(c) *Compliance with high court* variable reflects the degree to which the executive branch of government complies with important decisions of the high court with which it disagrees. The variable can take five distinctive values that have been converted to the interval using item response measurement model: (i) zero if the government never complies, (ii) one if the government seldom complies, (iii) two if the government complies about half of the time, (iv) three if the government usually complies, and (v) four if the government always complies with the important decisions of the high court. Higher values simply indicate a higher overall degree of compliance and respect of the judicial autonomy and decisions of the high court.

(d) *Judicial constraints* on the executive reflect and summarize the overall strength of the judiciary in constraining arbitrary government action that oftentimes violates the constitution and existing legislation. This particular variable is a high-level indicator of the extent to which the executive branch of government respects the constitution, complies with court rulings and the extent to which the judiciary is able to act independently of government pressure or threats. More specifically, the index is constructed through a point estimate from Bayesian factor analysis model of the following sub-level indicators: (i) the respect of the constitution by executive branch, (ii) compliance with judiciary, (iii) compliance with high court, (iv) high-court independence,

(v) lower-court independence. Higher values in the interval designate more rigorous and stronger judicial constraints on the executive branch of government.

(e) *Judicial purges* variable indicates the degree to which judges are removed from their posts for arbitrary and typically political reasons, and not for reasons such as strong evidence of corruption and abuse of judicial power. The variable can take five distinctive values: (i) zero if there was a massive and arbitrary purge of the judiciary, (ii) one if there were limited but very important arbitrary removals, (iii) two if there were limited arbitrary removals, (iv) three if judges were removed from office but there is no evidence that removals were arbitrary, and (v) four if judges were not removed from their posts. The variable is measured on the ordinal scale and converted to the interval through the item-response measurement model.

(f) *Judicial accountability* variable indicates the degree to which judges are removed from their posts or otherwise disciplined when found responsible for serious misconduct. The variable can take five different values: (i) zero if removal and disciplinary procedure never happen, (ii) one if the removal and disciplinary procedure seldom happen, (iii) two if judges are removed from their post or disciplined about half of the time, (iv) three if removal and disciplinary action usually happen, and (v) four if removal and disciplinary action always happen. Higher values of the variable indicate substantially greater degree of judicial accountability.

Figure 1 plots judicial outcomes of Türkiye against the rest of the world for the full period of investigation.[7]

---

[7] For the sake of convenience, Türkiye is highlighted black whilst the donor countries are highlighted in light grey

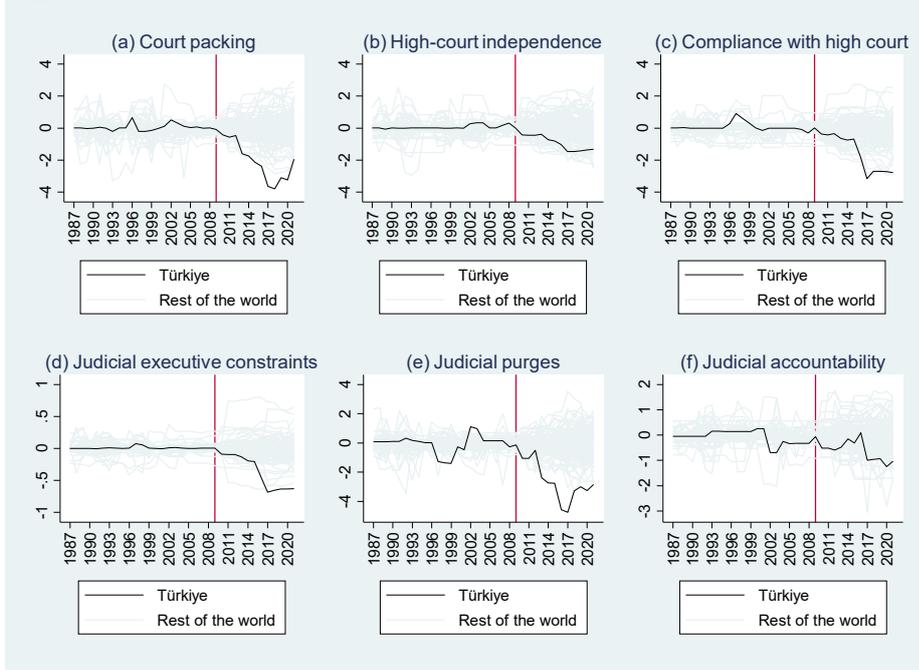

Figure 1: Judicial independence trends in Türkiye and the rest of the world, 1987-2023

### 4.1.2 Auxiliary covariates

The battery of the auxiliary covariates included into the **X** vector includes rich past judicial independence dynamics proxied by the values of the dependent variables in six different benchmark years[8] preceding the date of populist constitutional reforms. The second class of covariates comprises institutional characteristics to capture salience and comparability of Türkiye with the donor pool in terms of constitutional and political attributes such. These variables include Freedom House indices of the rule of law, political rights and civil liberties, the first principal component of Worldwide Governance Indicators (Kaufmann et. al. 2011), lexical index of electoral democracy (Skaaning et. al. 2015) and a latent measure of constitutional structure (Ginsburg 2012). These variables allow us to capture institutional and constitutional similarities between Türkiye and the donor pool through synthetic matching on observable characteristics. The third class of auxiliary covariates consists of the per capita GDP (Bolt et. al. 2018), ethnic and religious fractionalization indices (Alesina et. al. 2003), and share of Muslim population from *Pew Research Center*. These variables allow us to match Türkiye with the donor pool in terms of the level of economic development, demographic structure, ethno-linguistic composition and the overall institutional quality, and posit a reasonably strong leverage against the omitted variable bias in estimating the missing counterfactual scenario.

### 4.1.3 Sample

---

[8] 1987, 1990, 1995, 2000, 2005, and 2009.

Our period of investigation begins in 1987 and ends in 2021. By late 1980s, Türkiye has accomplished both political and economic liberalization and applied for the membership in the European Economic Community which provides a relatively stable and enduring time period without major overhauls and civil unrest that would pose a source of systemic instability. The overall sample comprises Türkiye as a single-treated country and a donor pool of 15 countries from the Mediterranean basin without armed conflicts[9] for the period of 37 years (i.e. 1987-2023) which yields a strongly balanced panel of 592 country-year paired observations.[10] Figure 2 presents the patterns of judicial outcomes by plotting the six outcome variables and the respective 95% confidence interval over time for Türkiye and the Mediterranean donor pool. For the sake of convenience, Türkiye is highlighted blue whilst the donor countries are highlighted in red. Although such analysis does not per se entail any inference on the effect, it serves as an informative insight into the overall outcome dynamics before a more nuanced and sophisticated analysis is undertaken. The raw trends of judicial independence imply a graphically visible deterioration of judicial independence with respect to increasing politically-motivated judicial appointments, lesser degree of independence and compliance, weak constraints and higher intensity of judicial purges after the populist constitutional reforms implemented in 2010. Further, judicial outcome trend in the donor pool is characterized by persistently stable dynamics without any major break in the underlying time series which ensure both salience and stability of the variation in judicial outcomes over time in safeguarding the isolation of the treatment effect of 2010 constitutional reforms.

---

[9] Albania, Algeria, Cyprus, Greece, Israel, Italy, Jordan, Malta, Mauritania, Morocco, North Macedonia, Portugal, Slovenia, Spain, and West Bank.

[10] It should be noted that our donor pool comprises only those countries from the Mediterranean donor pool which had not undergone political turmoil or armed conflicts such as Syria, Lebanon, Libya, Egypt and Tunisia as well as several countries from former Yugoslavia confronting prolonged armed conflict such as Croatia, Bosnia and Herzegovina and Serbia.

**Figure 2**: Judicial independence trajectories in Türkiye and Mediterranean donor states, 1987-2023

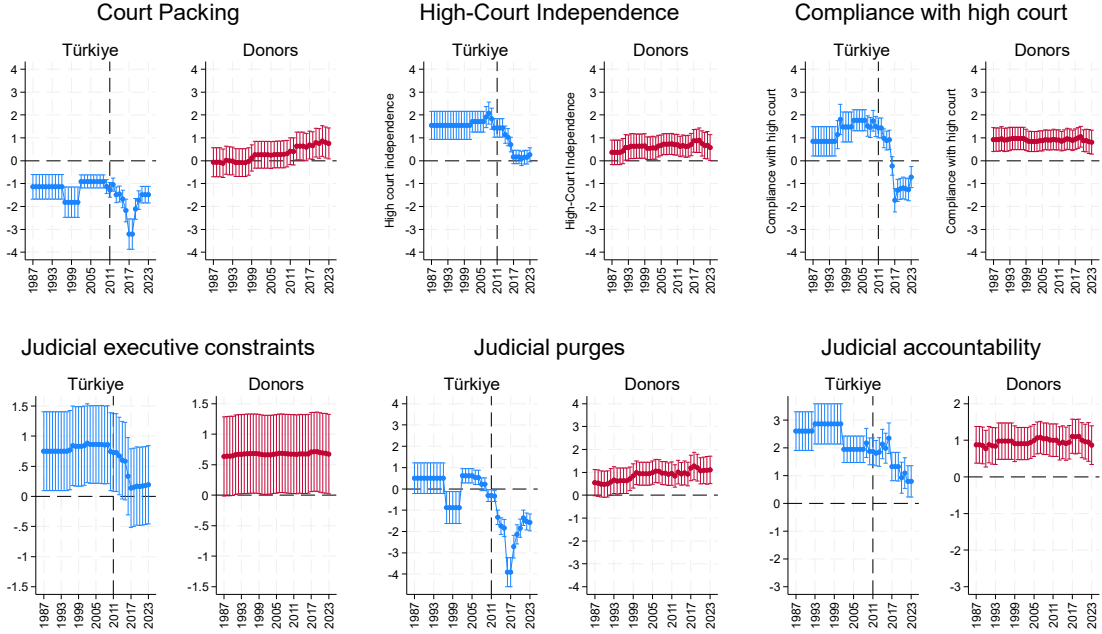

# 5 Results

## 5.1 Pre-reform outcome and auxiliary covariates balance

Table 1 draws out the judicial independence characteristics of Türkiye in the period before the constitutional reforms and compares them to those of the synthetic Türkiye constructed from a weighted combination of attributes of countries in the donor pool. Such preliminary analysis serves two distinctive purposes. First, the comparison can detect whether large discrepancies exist between the treated unit and its synthetic peer. Large discrepancies between the two usually indicate a poor quality of the fit and suggest that the synthetic control method is unable to provide any discernible evidence of the impact of policy intervention on the outcome of interest. And second, this particular comparison is also able to parse out some degree of confidence in the ability of the method to capture and reproduce the similarities between the treated unit and the synthetic control group. Although the size of RMSE depends on the scale of the dependent variable of interest, lower values relative to the mean and standard deviation typically indicate a better quality of fit. A close similarity of the outcome trajectory between the treated unit and its synthetic control group usually indicates a reasonably good ability of the method to quantitatively track and reproduce the outcome trajectory of the treated unit from the implicit attributes of the units in the donor pool.

In overall terms, the comparison in Table 1 reveals that the synthetic version of Türkiye provides a very good comparison for the actual Türkiye. In particular, the synthetic peer is very similar to the actual Türkiye with respect to the pre-2010 judicial independence outcomes, political and institutional characteristics, level of economic development as well as ethno-linguistic composition. In particular, the quality of the fit appears to be nearly excellent for the pre-reform judicial independence outcomes as the values of the actual Türkiye are completely on par with the counterpart values of its synthetic version indicating very low predictive discrepancy, and suggesting that other countries attributes are perfectible able to reproduce the judicial independence trajectories of Türkiye prior to the constitutional reforms with almost non-existing idiosyncrasies involved. Türkiye also matches well on the past outcomes' dynamics and the implicit auxiliary covariates with its synthetic counterpart. In particular, the synthetic versions of Türkiye appear to have a similar level of institutional quality and political rights and comparable level of civil liberties. Moreover, Panel C lays out the similarities with respect to the per capita GDP, population density and ethno-linguistic structure. It becomes apparent that the synthetic version of Türkiye has consistently similar level of per capita GDP, very similar ethno-linguistic fractionalization but somewhat higher value of the per capita GDP that the control group most plausibly reproducing and predicting the judicial independence trajectories of the synthetic version of Türkiye. For each outcome specification, the normalized variable weight $\hat{V}$ is reported. It becomes apparent that past judicial outcomes explain between 40% and 60% of the implied outcome variance which implies that additional predictors included in the specification cannot be considered irrelevant in achieving salience between Türkiye and the respective control groups. Evaluating the quality of the fit against the benchmark of the null model as proposed by Adhikari and Alm (2016), the RMSEs for Türkiye have less than 1% of the pre-treatment error margin which implies that the quality of the fit appears to be in the range between very good and excellent.

Table 2 reports the performance metrics associated with the application of synthetic control estimator to evaluate the effects of constitutional reforms on judicial outcomes under investigation. The evidence readily suggests that the synthetic control estimator under nested optimization route provides an excellent quality of the fit between Türkiye and its synthetic peer prior to the constitutional reforms. In particular, the imbalance between Türkiye and its synthetic version in the pre-treatment period decreases between 12% in court packing specification and 42% in high-court independence specification. In addition, post-synthetic matching bias decrease consecutively for each outcome under the investigation. The magnitude of bias reduction appears to be the largest in specification of court packing and judicial purges outcomes

Table 1: Covariate balancing between Türkiye and its synthetic control group before the constitutional reforms

| | Court Packing | | | High-Court Independence | | | Compliance with High Court | | | Judicial Constraints on Executive | | | Judicial Purges | | | Judicial Accountability | | |
|---|---|---|---|---|---|---|---|---|---|---|---|---|---|---|---|---|---|---|
| | Real Türkiye | Synthetic Türkiye | V Weight | Real Türkiye | Synthetic Türkiye | V Weight | Real Türkiye | Synthetic Türkiye | V Weight | Real Türkiye | Synthetic Türkiye | V Weight | Real Türkiye | Synthetic Türkiye | V Weight | Real Türkiye | Synthetic Türkiye | V Weight |
| RMSE | | 0.288 | | | 0.091 | | | 0.282 | | | 0.031 | | | 0.525 | | | 0.374 | |
| **Panel A: Past outcome dynamics covariates** | | | | | | | | | | | | | | | | | | |
| Outcome level in 1987 | -1.138 | -1.133 | 0.119 | 1.547 | 1.534 | 0.087 | 0.849 | 0.943 | 0.153 | 0.750 | 0.738 | 0.121 | 0.504 | 0.440 | 0.262 | 2.600 | 2.437 | 0.072 |
| Outcome level in 1990 | -1.138 | -1.133 | 0.211 | 1.547 | 1.534 | 0.028 | 0.849 | 1.117 | 0.100 | 0.750 | 0.757 | 0.152 | 0.504 | 0.440 | 0.256 | 2.600 | 2.437 | 0.118 |
| Outcome level in 1995 | -1.138 | -1.469 | 0.004 | 1.547 | 1.538 | 0.157 | 0.849 | 1.305 | 0.034 | 0.750 | 0.789 | 0.117 | 0.504 | 0.074 | 0.062 | 2.857 | 2.437 | 0.189 |
| Outcome level in 2000 | -1.816 | -1.151 | 0.033 | 1.547 | 1.624 | 0.213 | 1.479 | 1.305 | 0.129 | 0.832 | 0.810 | 0.246 | -0.882 | 0.074 | 0.141 | 2.857 | 2.367 | 0.046 |
| Outcome level in 2005 | -0.910 | -1.237 | 0.000 | 1.709 | 1.744 | 0.175 | 1.767 | 1.537 | 0.057 | 0.863 | 0.851 | 0.281 | 0.617 | 0.221 | 0.052 | 1.942 | 2.367 | 0.031 |
| Outcome level in 2009 | -0.910 | -1.007 | 0.131 | 1.835 | 1.733 | 0.162 | 1.744 | 1.537 | 0.250 | 0.858 | 0.851 | 0.061 | 0.217 | 0.226 | 0.063 | 2.173 | 2.350 | 0.015 |
| **Panel B: Institutional covariates** | | | | | | | | | | | | | | | | | | |
| Rule of Law | 8 | 7.92 | 0.076 | 8. | 13.82 | 0.001 | 8 | 12.65 | 0.001 | 8 | 13.86 | 0.000 | 8.00 | 6.59 | 0.050 | 8 | 10.26 | 0.014 |
| Political Rights | 3.26 | 4.30 | 0.031 | 3.25 | 1.25 | 0.001 | 3.26 | 2.03 | 0.063 | 3.26 | 1.56 | 0.001 | 3.26 | 4.46 | 0.015 | 3.26 | 1.39 | 0.028 |
| Civil Liberties | 4.04 | 4.26 | 0.069 | 4.00 | 1.57 | 0.001 | 4.04 | 2.23 | 0.001 | 4.04 | 1.69 | 0.001 | 4.04 | 4.88 | 0.016 | 4.04 | 2.52 | 0.011 |
| Quality of Governance (1996-2010) | -1.061 | -0.999 | 0.227 | -1.06 | 2.09 | 0.001 | -1.06 | 1.35 | 0.005 | -1.06 | 1.86 | 0.00 | -1.06 | -1.06 | 1.31 | -1.06 | 1.31 | 0.03 |
| Lexical Index of Democracy (1987-2010) | 6 | 2.669 | 0.004 | 6.00 | 5.93 | 0.016 | 6 | 4.80 | 0.046 | 6 | 5.56 | 0.003 | 6 | 6 | 6 | 6 | 6 | 0.271 |
| Constitutional Structure (1987-2010) | -1.173 | 0.025 | 0.001 | -1.17 | -0.98 | 0.095 | -1.17 | -0.75 | 0.106 | -1.17 | -0.99 | 0.01 | -1.17 | -1.17 | -0.10 | -1.17 | -0.10 | 0.15 |
| **Panel C: Auxiliary covariates** | | | | | | | | | | | | | | | | | | |
| Ethnic fractionalization | 0.32 | 0.32 | 0.014 | 0.320 | 0.125 | 0.000 | 0.320 | 0.093 | 0.002 | 0.320 | 0.104 | 0.000 | 0.320 | 0.156 | 0.004 | 0.320 | 0.353 | 0.014 |
| Linguistic fractionalization | 0.22 | 0.22 | 0.001 | 0.221 | 0.225 | 0.066 | 0.221 | 0.236 | 0.035 | 0.221 | 0.313 | 0.002 | 0.221 | 0.144 | 0.017 | 0.221 | 0.549 | 0.005 |
| Share Muslim | 0.94 | 0.79 | 0.039 | 0.935 | 0.156 | 0.000 | 0.935 | 0.364 | 0.019 | 0.935 | 0.303 | 0.004 | 0.935 | 0.800 | 0.037 | 0.935 | 0.190 | 0.005 |
| Per Capita GDP | 11400 | 8549 | 0.007 | 11613 | 19908 | 0.000 | 11400 | 17075 | 0.000 | 11400 | 18285 | 0.001 | 11400 | 10059 | 0.014 | 11400 | 22933 | 0.006 |

Table 2: Performance of synthetic control estimator

|  | Court Packing | High-Court Independence | Compliance with High Court | Judicial Constraints on Executive | Judicial Purges | Judicial Accountability |
|---|---|---|---|---|---|---|
| RMSE (with nested optimization) | 0.288 | 0.091 | 0.282 | 0.031 | 0.525 | 0.374 |
| RMSE (without nested optimization) | 0.326 | 0.155 | 0.376 | 0.041 | 0.603 | 0.514 |
| Performance improvement | 12% | 42% | 25% | 25% | 13% | 27% |
| Bias without synthetic matching | 0.58 | 0.42 | 0.23 | 0.24 | 0.27 | 0.40 |
| Bias with synthetic matching | 0.07 | 0.24 | 0.20 | 0.23 | 0.03 | 0.13 |
| Implied bias reduction | 88% | 43% | 14% | 3% | 88% | 67% |
| # control units | 16 | 16 | 16 | 16 | 16 | 16 |
| # covariates and pre-overhaul outcomes | 17 | 17 | 17 | 17 | 17 | 17 |

Notes: the table reports the imbalance between Türkiye and its synthetic control group across the full set of outcome specification approximated through the root mean squared prediction error (RMSE) with and without the nested optimization route and the estimated bias with and without synthetic matching scheme, and the implicit bias reduction after the application of synthetic matching algorithm to address potential discrepancies between Türkiye and its donor pool in the pre-intervention period.

### 5.2   Composition of synthetic control group

Table 3 reports a detailed composition of Türkiye's synthetic control groups for each judicial independence outcome. The composition of control groups unveils whether the set of convex characteristics and attributes of countries prior to the constitutional referendum can quantitatively track and reproduce the respective outcome trajectory. Provided that the intervention of interest is reasonably unique and not perceptible elsewhere in the donor pool, the convex combination of implied attributed constructed the optimal and best-fitting non-zero weights tend to approximate the most plausible and likely counterfactual scenario in the hypothetical absence of the intervention.

Our evidence reveals substantial similarities between Türkiye and the Mediterranean donor countries in the period before the constitutional referendum. For instance, Türkiye's trajectory of court packing prior to 2010 can be best reproduced as the convex combination of explicit and implicit attributes West Bank (26%), Mauritania (26%), Cyprus (19%), Jordan (15%), and Malta (6%). The respective vector of weights implied that these countries best reproduce Türkiye's court packing characteristics prior to the reforms. Notice that the explicit similarity emanates from the similarity of pre-reform outcomes whilst the implicit one arises from the similarity in terms of the auxiliary attributes such as political rights, rule of law, civil liberties, ethno-linguistic composition, level of economic development and constitutional structure.

In terms of further example, Türkiye's trajectory of high-court independence in 1987-2010 period can be best reproduce through the weighted average of the implied attributes of Cyprus (60%), Portugal (28%), and North Macedonia (12%), respectively. In the specification using the degree of compliance with high court as the outcome, Türkiye's synthetic control group is dominated by Cyprus (80%) followed by a set of countries with relatively lower weight shares such as West Bank (26%) and Portugal (24%). Furthermore, Türkiye's trajectory of judicial constraints on the executive prior to the constitutional reforms is best reproduced through the convex combination of the full-fledged attributes of Cyprus (75%) followed by a relatively more minor influence of Albania (9%), Portugal (9%), and West Bank (6%). In addition, the synthetic control group in the judicial purges' specification appears to be equally concentrated and less dense. It consists of West Bank (76%), Italy (20%), and Mauritania (4%). Moreover, the composition of control group in the judicial accountability specification is even more dense and consists of only two donor countries most capable of reproducing Türkiye's trajectory of judicial accountability before the populist constitutional reforms, namely, Israel (93%) and North Macedonia (7%), respectively.

Regardless of the compositional differences behind the multiple-outcome synthetic control groups, the most plausible question to ask is whether the countries entering the control groups with non-zero weight and thereby contributing to the counterfactual outcome trajectory are tainted by the presence of populist constitutional reforms in a similar way as Türkiye. Since none of the countries in the synthetic control groups has undergone populist constitutional backsliding prior to 2010, it is highly unlikely that the synthetic control groups would be subject to fragile internal validity quality or to the violation of stable unit treatment value assumption. Countries prone to the violation of these two criteria should be discarded from the donor pool prior to the conduct of the synthetic control analysis.

Table 3: Composition of Türkiye's synthetic control groups

|  | Court Packing | High-Court Independence | Compliance with High Court | Judicial Constraints on Executive | Judicial Purges | Judicial Accountability |
|---|---|---|---|---|---|---|
| Albania | 0 | 0 | 0 | 0.09 | 0 | 0 |
| Algeria | 0 | 0 | 0 | 0 | 0 | 0 |
| Cyprus | 0.19 | 0.60 | 0.50 | 0.75 | 0 | 0 |
| Greece | 0 | 0 | 0 | 0 | 0 | 0 |
| Israel | 0 | 0 | 0 | 0 | 0 | 0.93 |
| Italy | 0 | 0 | 0 | 0 | 0.20 | 0 |
| Jordan | 0.15 | 0 | 0 | 0 | 0 | 0 |
| Malta | 0.06 | 0 | 0 | 0 | 0 | 0 |
| Mauritania | 0.26 | 0 | 0 | 0 | 0.04 | 0 |
| Morocco | 0 | 0 | 0 | 0 | 0 | 0 |
| North Macedonia | 0 | 0.12 | 0 | 0 | 0 | 0.07 |
| Portugal | 0 | 0.28 | 0.26 | 0.09 | 0 | 0 |
| Slovenia | 0 | 0 | 0 | 0 | 0 | 0 |
| Spain | 0 | 0 | 0 | 0 | 0 | 0 |
| West Bank | 0.33 | 0 | 0.24 | 0.06 | 0.76 | 0 |

## 5.3 Baseline results

Figure 3 reports the effect of 2010 constitutional reforms on judicial independence in Türkiye for the period 1987-2021, estimated through the application of the synthetic control method. The figure depicts the evolution of the actual outcome and the estimated counterfactual scenario indicated by the synthetic Türkiye in the hypothetical absence of the constitutional reforms. Provided that the quality of fit is reasonably good, the comparison between the actual Türkiye and its synthetic peer allows for the impact of constitutional reforms to be evaluated. Judged by the size of pre-2010 root mean square prediction error (RMSE) relative to the scale of outcome variable, our analysis suggests that the synthetic control groups provide a very good characterization and approximation of Türkiye prior to the reforms through past outcome dynamics and auxiliary attributes.

The evidence indicates that the 2010 constitutional reforms are closely linked to a widespread deterioration in judicial independence. Notably, the actual outcomes following 2010 consistently fall short of the estimated counterfactuals. The erosion of judicial independence, driven by populist constitutional backsliding, is evident in several areas: a marked increase in politically motivated judicial appointments to both higher and lower courts, a weakening of high court independence under government pressure, more frequent instances of government non-compliance with court rulings, the erosion of executive constraints, and a significant rise in the arbitrary removal of judges, typically for political reasons. The gap between the actual and counterfactual outcomes underscores the severity of the decline. The estimated counterfactual trajectories point to a scenario in which judicial independence would have remained much stronger, characterized by greater resilience against government interference and the absence of systematic attacks on the judiciary. This erosion of judicial independence, as revealed by the discrepancy between actual and counterfactual scenarios, unfolded rapidly and has shown persistent and enduring effects.

The negative effect of populist constitutional backsliding appears to increase substantially over time and is further underpinned by its permanency. From the first year of the constitutional backsliding up to the present day, the court packing index decreases by 2.18 basis points or 1.65 times the standard deviation, high-court independence index drops by 1.20 basis points (=0.9 times the standard deviation), index of government's compliance with the high court deteriorates by 2.6 basis points (=2.7 times the standard deviation), judicial constraints index deteriorates by 0.57 basis points (=twice the standard deviation) whereas the index of judicial purges decreases by 2.87 basis points (=2.8 times standard deviation) relative to the synthetic control group. The only component of judicial independence that appears to be unaffected by the populist constitutional reforms appears to be judicial accountability whereupon the break between the actual and synthetic version of the index is not perceptible and the present-day gap does not seem to be attributed to the constitutional reforms initiated in 2010. Since the estimated deterioration of judicial independence and related outcome exceeds the standard deviation over the full period of investigation, the magnitude of the effect of constitutional reform does not seem to be small.

**Figure 3**: Effects of constitutional reform on judicial independence in Türkiye, 1987-2021

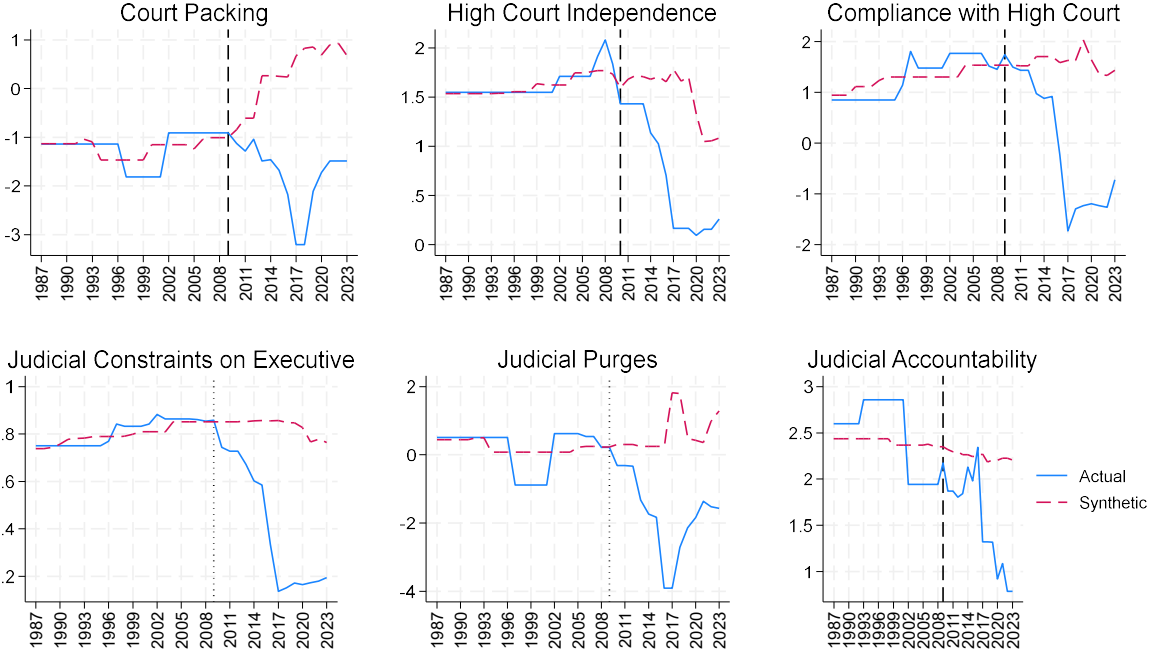

Our synthetic control estimates highlight a dramatic and irreversible erosion of judicial independence triggered by the populist constitutional reforms, which encouraged backsliding rather than promoting the improvement of judicial quality and independence. The large and persistent negative gaps between Türkiye and its synthetic control group reinforce the idea of a permanent breakdown in judicial independence, manifested in the intensification of politically motivated judicial appointments and purges, the weakening of judicial constraints, and the government's increasing disregard for judicial authority. As these negative gaps endure over time, our findings contradict any suggestion that the populist backlash against the independent judiciary—as a check on executive power—might have a temporary or transient effect. Instead, the collapse of judicial independence appears to be permanent, underscoring the irreversible damage inflicted by these constitutional reforms. Although there is some indication of recovery in the trajectory of judicial purges, where the intensity of the effects diminishes slightly over time, the deep and significant negative consequences of the populist reforms persist. The key question, then, is whether the estimated impacts of these populist reforms on judicial independence are robust to various specification checks and statistically significant at conventional levels. Our evidence suggests they are, further confirming the long-term repercussions of Türkiye's populist constitutional reforms.

*5.4. In-time placebo analysis*

One of the most obvious caveats behind the inference in small-sample comparative case studies concerns the credibility of the estimated treatment effect. In particular, the credibility of the effect of 2011 constitutional reforms on judicial independence in Türkiye could be jeopardized if the deterioration in compliance and independence of the judiciary if the estimated gaps were similarly large if the 2011 referendum were reassigned to the year other than 2011. Under such circumstances, confidence in the validity of the results would diminish if the synthetic control method estimated similarly large impacts when applied to deliberately false dates when the constitutional referendum did not occur.

Against this backdrop, two distinct challenges emerge as potential threats to the credibility of the estimated gaps. First, the decline in judicial independence might have been set in motion by the broader democratic backsliding triggered by the Justice and Development Party's (AKP) landslide electoral victory in 2002. In this scenario, shifting the treatment year to 2002 would show a significant divergence between Türkiye and its synthetic counterpart starting with the AKP's electoral win, rather than the 2010 constitutional referendum. Second, in 2017, another series of constitutional reforms was implemented following a referendum on 21 proposed amendments. This referendum took place under emergency rule following the failed coup attempt of 2016 (Gökmenoğlu 2024). The amendments, approved by 51% of the vote, further eroded checks and balances by consolidating executive power. If the estimated gaps in judicial independence are truly a consequence of the 2010 reforms, reassigning the treatment date to 2017 should reveal that the divergence between synthetic and actual Türkiye began after 2010 and did not arise in response to the 2017 constitutional amendments.

As a falsification analysis, we conduct both in-time placebo analyses through two-stage permutation of the treatment to the deliberately wrong dates of the 2011 constitutional referendum. First, through backward permutation of the 2011 referendum to the year 2002, our approach can detect whether the erosion of judicial independence and constraints on the executive has been primarily triggered by the democratic backsliding commencing with the landslide electoral victory of AKP party in the ensuing year. By contrast, through the forward permutation of 2011 referendum to the year 2017, we can examine whether the constitutional amendments approved in 2017 referendum posit a differential and distinctive shock to the validity and credibility of the estimated counterfactual scenario.

Figure 4 reports the result of the backward in-time placebo analysis of 2011 referendum. It becomes apparent that the synthetic Türkiye almost exactly replicates the trajectory of judicial outcomes in the actual Türkiye for the 1987-2002 period. Perhaps most importantly, the trajectories of judicial outcomes fail to diverge considerably during the 1987-2002 period. Instead, the divergence between the actual Türkiye and its synthetic peer across all six judicial outcomes unfolds immediately after 2011. Contrary to the actual constitutional referendum in 2011, the placebo date of the referendum has no perceivable effect on judicial outcomes whatsoever. In turn, the evidence readily suggests that our baseline results reflect the impact of the 2011 constitutional referendum on judicial independence from government pressures instead of the broader and more profound effect of democratic backsliding.

**Figure 4**: Backward permutation of constitutional referendum to deliberately false date, 1987-2023

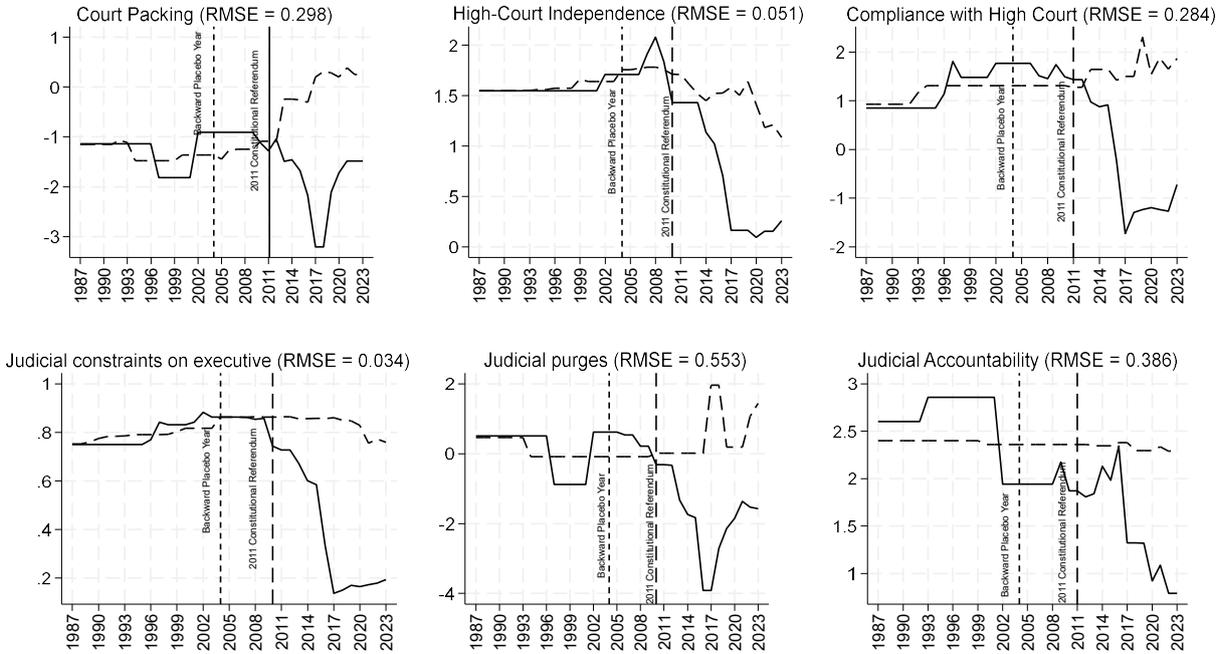

Figure 5 reports the result of the forward permutation of the constitutional referendum to the year 2017. The results of the in-time treatment reassignment suggest that the divergence between synthetic and real Türkiye fails to unfold in 2017. In particular, the trajectories of judicial outcomes between the actual and synthetic Türkiye moves in tandem with very little discrepancy in the period 1987-2011 whereas the trajectories cannot be synthetically matched with an equally low discrepancy in the period 2011-2017. Instead, a strong indication is posited for the structural break in judicial independence and the related trajectories taking place in 2011. Despite a notable reduction of the intensity of judicial purges and court packing, the synthetic control estimator fails to match Türkiye and its synthetic Mediterranean control group for the period 1987-2017 whilst the evidence only reinforces our prior results indicating a substantial structural break unleashed by the constitutional referendum in 2011. In addition, the reversal of court packing and judicial purges should not be viewed as surprising given that the process of domesticating and subordinating the judicial branch of government to the executive power has been largely completed after 2017 while our results indicate that the process of the government-orchestrated assault on the judiciary has not been unleashed in 2017 but in 2011 after the constitutional referendum served as a formal door-opener to the process of judicial backsliding as well as a window-dressing scheme to bolster a popular approval for such large-scale attack on judicial independence and integrity. Furthermore, our results from the forward treatment permutation almost unequivocally indicate that the divergence between synthetic and real Türkiye in high-court independence, compliance with the high court and judicial constraints on the executive appears to be permanent up to the present day with almost no evidence of reversal of temporariness of the effect at hand.

**Figure 5**: Forward permutation of constitutional referendum to deliberately false date, 1987-2023

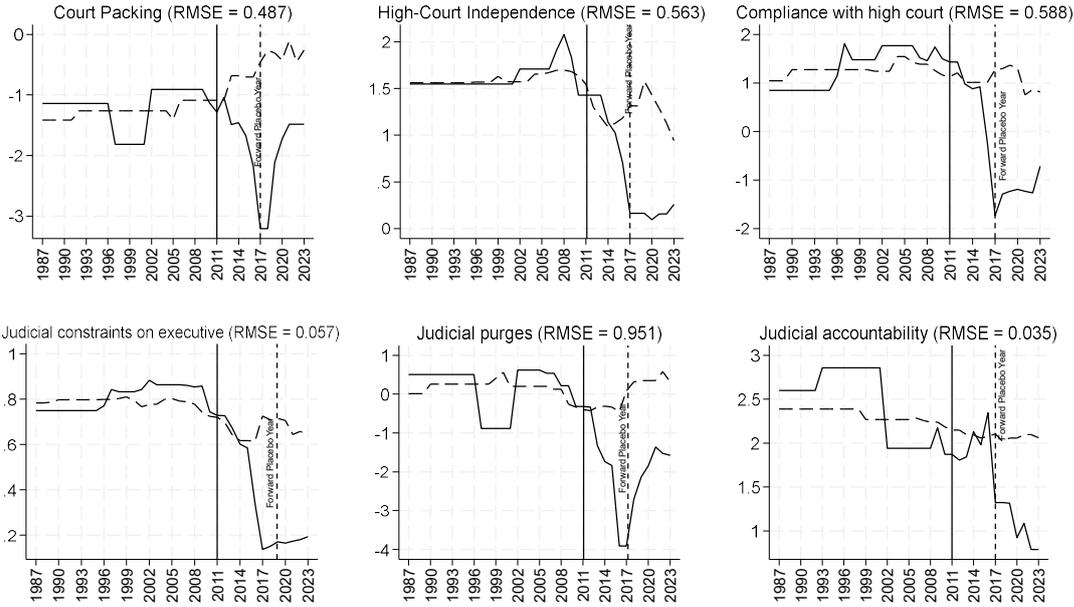

### 5.5  In-space placebo analysis

To tackle the statistical significance of the estimated judicial independence gaps between Türkiye and its control groups, we ask whether the estimated gaps could be driven almost entirely by chance alone. Recall that the in-space placebo analysis is conducted through an iterative assignment of the populist constitutional reforms to the other countries that have not undergone Erdoğan-style constitutional backsliding and the subsequent erosion of judicial independence.

This allows us to determine whether the negative judicial independence gaps of such magnitude would be obtained if we had chosen a country randomly instead of Türkiye. Hence, by applying the synthetic control estimator to the countries that did not implement the specific set of independence-eroding constitutional reforms of the judiciary, the placebo simulation can be run. The notion behind the placebo analysis is both intuitive and straightforward. If the placebo simulation yields judicial independence gaps of very similar magnitude to the ones estimated for Türkiye, the subsequent interpretation is that the analysis does not provide any significant evidence of the negative impacts of the populist constitutional backsliding since similar gaps can be equally likely perceptible elsewhere. By contrast, if the placebo simulation uncovers gaps for Türkiye being both relatively unique and unusually large compared to the countries without the respective reforms, the notion of significant evidence of the negative effect of populist constitutional reforms becomes more plausible and credible. The placebo analysis is carried out in two steps. In the first step, the countries in the donor pool are iteratively assigned the constitutional reform-related treatment in 2011 which moves Türkiye to the donor pool. In the next step, the estimated impact is computed for each country and each respective outcome of interest which yields a series of distributions of the estimated gaps for the countries where no intervention took place. Finally, the proportion of countries with the estimated impacts as large as the ones estimated for Türkiye is calculated for each year in the post-intervention period. In addition, the estimated judicial independence gaps for Türkiye are also examined through a parametric difference-in-difference analysis to determine their respective uniqueness and significance compared to the placebo gaps. Whilst the first part of the placebo analysis is non-parametric, the second part proceeds with a parametric analysis relying on parallel trends in the gaps between Türkiye and the corresponding placebos. The latter does not indicate the fraction of countries with gaps similar to Türkiye but is able to uncover whether the treatment-related gaps can be properly and statistically distinguished from the placebo gaps in the full post-intervention period and is somewhat less susceptible to the distributional assumption behind such analysis.

Figure 6 reports the differential between post-2010 and pre-2010 RMSE for Türkiye and the full set of countries not undergoing the designated intervention. The evidence clearly suggests that the negative gaps for Türkiye appear to be somewhat unique and not perceptible elsewhere in the donor pool. The negative gaps are particularly large and pronounced for the outcomes pertaining to court packing, compliance, judicial constraints and judicial purges. The gap appears to moderately large for high-court independence whilst it seems to be weak for judicial accountability-related outcome. Figure 6 also indirectly propounds that the synthetic control method provides a very good fit for judicial independence trajectories of Türkiye in the years before the populist dismantling of the judiciary in 2010. If the synthetic versions of Türkiye were somehow unable to fit the judicial independence trajectories of real Türkiye in the years prior to the constitutional referendum, it would become obvious that that post-intervention gap between the actual and synthetic version for Türkiye is fabricated by the lack of fit rather than reflecting the effect of constitutional reforms. To provide some information about the relative rarity of estimating a large post-reform gaps for countries that were well fitted before the reforms, we follow Abadie et. al. (2010) and discard those entities from the donor pool having a pre-reform mean square prediction error more than four times the error for Türkiye to ensure that the latter is compared only to those countries fitting almost as well as Türkiye in the pre-intervention period.

Figure 6: Uniqueness of the effect of 2010 constitutional referendum on judicial independence

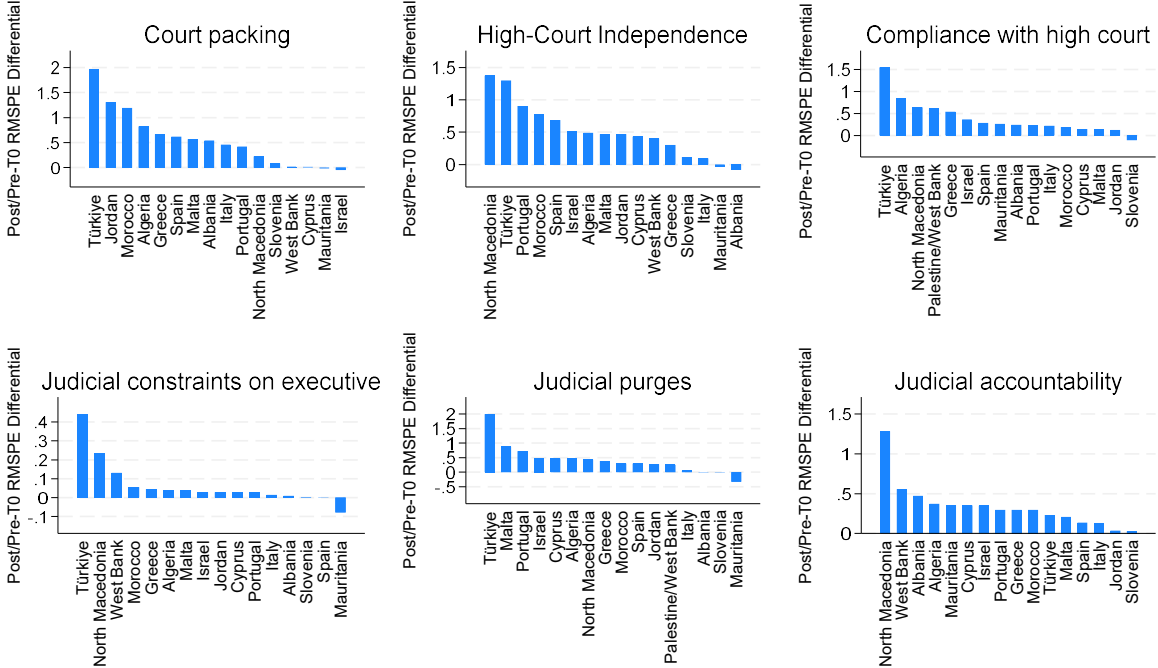

In Figure 7, we further evaluate the estimated gaps for Türkiye relative to the gaps obtained in the placebo simulation by looking at the distribution of the post/pre-reform ratios of RMSE. Instead of reporting a distribution, we compute the proportion of countries having the estimated gaps as high as Türkiye which can be roughly interpreted as a probability that Türkiye's gaps would happen at random. Notice that higher values of the proportion indicate somewhat greater chances of obtaining the effects at random. The evidence suggests that the approximated probabilities are consistently low and within the 10% probability bound for the entire post-intervention period for court packing, judicial constraints on the executive and judicial purges. That said, the negative gap between Türkiye and its synthetic control group for these outcomes tends to unfold immediately and is underpinned by its permanency. The estimated gaps for high-court independence and compliance do not appear to be statistically significant in the first few years after the intervention whilst the probabilities of random effect gradually rescind towards zero up to the present day, suggesting a permanent but more gradual erosion of the judicial independence and government's compliance with the judiciary. By contrast, the probabilities for judicial accountability are high in the short period after the intervention but over time also exhibit a tendency towards the conventional statistical significance at the usual bounds.

**Figure 7**: Inference on the effect of constitutional reforms

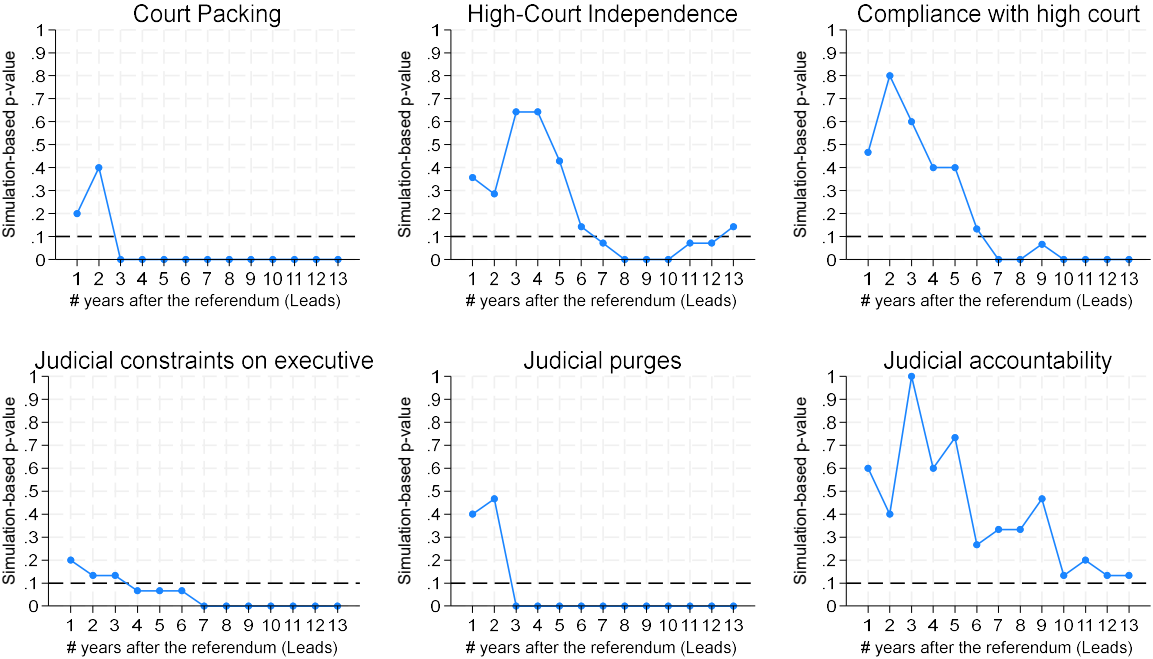

Table 4 reports the difference-in-differences estimated gaps comparison between Türkiye and the gaps obtained through the placebo runs. The key parameter of interest reflects the post-intervention magnitude of the coefficient on the average treatment effect for Türkiye in comparison with the placebo gaps which allows for a parametric analysis of the gaps to unravel the possible statistical significance. The evidence suggests that post-intervention outcome gaps for Türkiye are both large, negative and statistically significant at 1% respectively. The coefficients on the estimated judicial independence gaps are consistently negative and particularly large for the court-packing and judicial purges variables whilst being noticeably smaller for the compliance, judicial constraints, accountability and high-court independence outcome variables. In each difference-in-differences specification, we also include two lags of the treatment and placebo-related gaps to control for the potential outcome persistence over time, and also add the full battery of both country-fixed effects and time-fixed effects to parse out the unobserved heterogeneity as a possible source of confounding influence as well as to ensure a reasonably robust difference-in-differences coefficient. In the lieu of mutually consistent specification, the parametric analysis of the gaps confirms the uniqueness for Türkiye relative to the placebo distribution.

Table 4: Parametric inference on the effect of constitutional reforms

|  | Court packing | High-court independence | Compliance with high court | Judicial constraints on the executive | Judicial purges | Judicial accountability |
|---|---|---|---|---|---|---|
|  | (1) | (2) | (3) | (4) | (5) | (6) |
| Effect | -.679*** | -.266*** | -.423*** | -.100*** | -.923*** | -.137*** |
|  | (.062) | (.062) | (.028) | (.006) | (.077) | (.014) |
| Two-tailed 95% Confidence bound | [-.803, -.556] | [-.318, -.214] | [-.478, -.367] | [-.113, -.087] | [-1.076, -.771] | [-.165, -.109] |
| Effect persistence (p-value) | (0.000) | (0.000) | (0.000) | (0.000) | (0.000) | (0.000) |
| # observations | 4,422 | 4,422 | 4,422 | 4,422 | 4,422 | 4,422 |
| # regions | 134 | 134 | 134 | 134 | 134 | 134 |
| Within R2 | 0.51 | 0.58 | 0.69 | 0.70 | 0.43 | 0.51 |
| Between R2 | 0.96 | 0.98 | 0.99 | 0.98 | 0.95 | 0.98 |
| Overall R2 | 0.57 | 0.64 | 0.75 | 0.75 | 0.51 | 0.59 |
| # Random donor samples | 1,000,000 | 1,000,000 | 1,000,000 | 1,000,000 | 1,000,000 | 1,000,000 |
| Permutation method | Random sampling | Random sampling | Random sampling | Random sampling | Random sampling | Random sampling |
| Country-fixed effects | YES | YES | YES | YES | YES | YES |
| (p-value) | (0.000) | (0.000) | (0.000) | (0.000) | (0.000) | (0.000) |
| Time-fixed effects | YES | YES | YES | YES | YES | YES |
| (p-value) | (0.415) | (0.502) | (0.183) | (0.531) | (0.705) | (0.572) |

Notes: the table reports the post-intervention difference-in-differences coefficients associated with judicial independence gaps after the constitutional reforms in 2011. In each specification, the full set of country-fixed effects and time-fixed effects is included. Standard errors of the actual and placebo gap coefficients are adjusted for arbitrary heteroscedasticity and serially correlated stochastic disturbances using finite-sample adjustment of the empirical distribution function with the error component model. Cluster-specific standard errors are denoted in the parentheses. Asterisks denote statistically significant coefficients at 10% (*), 5% (**), and 1% (***), respectively.

### 5.6 Differential trend analysis

Figure 8 depicts differential trend analysis of the impact of 2011 constitutional reforms. It compares the time trend in the Türkiye's judicial independence gap relative to its synthetic control group with the time trend of the gap in the post-intervention period. The purpose of such analysis is to unravel whether the constitutional reforms induced a differential trend in the judicial independence outcomes. The comparison of the change in the slop of actual Türkiye before and after the constitutional reforms with the change in the slope of its synthetic peer before and after the constitutional reforms allows to determine how much larger was the change in the erosion of judicial independence in actual Türkiye and its synthetic counterpart before vs. after the reforms. Moreover, it should be noted that such comparison also allows us to investigate whether the constitutional reforms induced a structural break in the trend of each respective outcome. The figure reports the slope of the gap's time trend in pre- and post-reform period alongside the local supremum $\tau$-statistics and the corresponding p-value. A failure to reject the null hypothesis would indicate a mere absence of the differential trend induced by the constitutional reforms. The evidence almost unequivocally suggests that the null hypothesis can be easily rejected at the conventional 5% bound for four specific outcomes: (i) court packing, (ii) high-court independence, (iii) compliance with high court, and (iv) judicial purges. Without the loss of generality, the evidence lends considerable empirical support in favor of the argument that the backbone of the populist constitutional backlash against judicial independence undertaken by Erdoğan's reforms consisted of politically-motivated judicial appointments, erosion of court's independence, undermined compliance and the arbitrary removals of judges from their posts either for arbitrary or politically-motivated reasons. By contrast, the null hypothesis of differential trends cannot be rejected for the judicial accountability (i.e. p-value = 0.904) where a visually clear downward time trend is perceivable, suggesting that pre-existing tendencies may be somewhat attributed to the estimate post-intervention gap. Furthermore, the null hypothesis on the differential trends can reject for the other five outcomes indicating a reasonably strong break in the time trend induced by the populist constitutional backsliding.

Figure 8: Testing differential trend assumption in response to populist constitutional reforms in Türkiye, 1987-2021

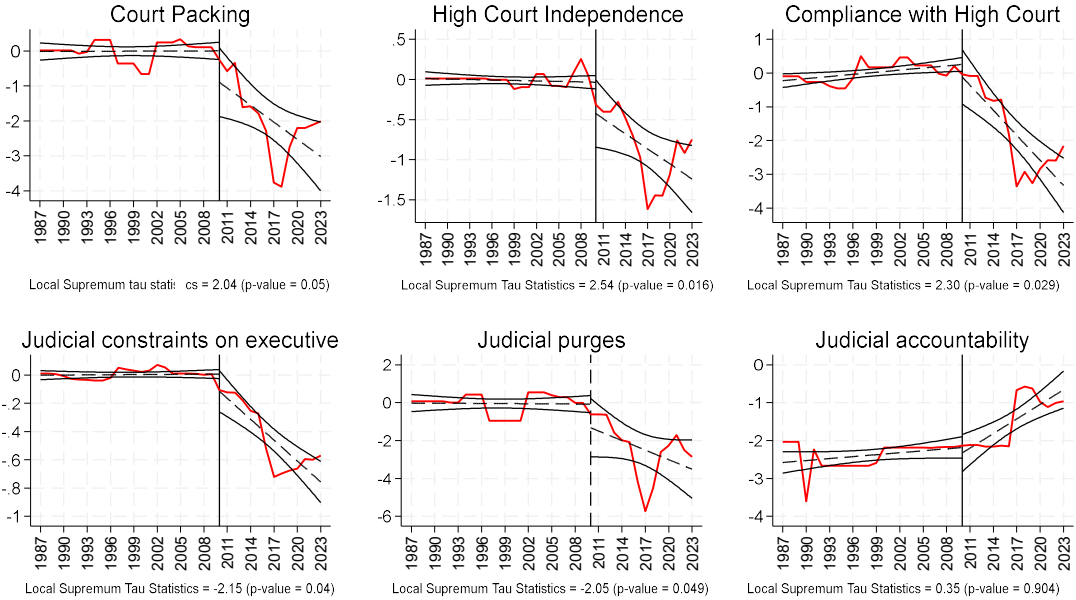

We assess the significance of the differential trend assumption further by performing additional structural break tests on the outcome-specific gap between real Türkiye and its synthetic peer to examine whether distinctive breakpoints are perceptible. Table 5 reports the results of the structural break tests in greater detail. In Panel A, we reported the results of simple structural break test for each specific outcome under known breakpoint. In the simplest possible terms, the null hypothesis on the judicial outcome variance and trend before and after 2011. We reject the null hypothesis at 5% significance threshold and show that 2011 constitutional reforms ushered fundamental structural changes to the judicial system both in terms of court packing, judicial constraints as well as compliance and independence from the executive interference. Consistent with our prior results, the null hypothesis on judicial accountability gap cannot be rejected even at artificially high significance threshold. Structural break in the trajectories of judicial independence and compliance with the high court is further reinforced by Chow test of structural break in Panel B. Furthermore, Panel C and Panel D exhibits the results of the residual-based cointegration test, allowing for the possibility of regime- and trend shifts (Gregory and Hansen 1996). Under the null hypothesis, no shift in the regime or trend of the outcome variable is evident. One of the notable advantages of the Gregory-Hansen test emanates from the identification of structural break date. The null hypothesis of no shift in trend and the regime of the variable is easily rejected at 1% significance threshold, respectively. Based on the test statistics behind the break, we find that structural change in the trajectory of judicial independence and other outcomes takes place between 2010 and 2013 which almost exactly coincides with the timing of the constitutional referendum. In essence, the results of the break tests unequivocally and almost undeniably confirm pervasive deterioration of judicial independence and compliance as well as a widespread escalation of court packing and judicial purges after the constitutional referendum, and suggest that the structural breakdown of the independence, compliance and constraints-related trajectories has not vanished up to the end-of-sample period, and, therefore, appears to be permanent.

Table 5: Testing for structural breaks in Türkiye's judiciary, 1987-2022

|  | Court Packing | High-Court Independence | Compliance with High Court | Judicial Constraints on Executive | Judicial Purges | Judicial Accountability |
|---|---|---|---|---|---|---|
| Panel A: Wald F-test for structural break with known breakpoint | | | | | | |
| F test statistics (p-value) | 4.18 (0.049) | 6.47 (0.016) | 5.29 (0.028) | 4.63 (0.045) | 4.21 (0.049) | 0.01 (0.906) |
| Panel B: Chow test for structural break | | | | | | |
| Fisher test statistics (p-value) | 5.62 (0.000) | 3.69 (0.016) | 5.64 (0.000) | 2.52 (0.064) | 5.62 (0.000) | 2.77 (0.047) |
| Panel C: Gregory-Hansen Test for Cointegration with Regime Shifts | | | | | | |
| Phillips $Z_t$ test statistics (p-value) | -6.63 (0.000) | -8.32 (0.000) | -8.15 (0.000) | -7.40 (0.000) | -7.24 (0.000) | -11.13 (0.000) |
| Estimated Break year | 2012 | 2011 | 2011 | 2011 | 2011 | 2013 |
| Panel C: Gregory-Hansen Test for Cointegration with Trend Shifts | | | | | | |
| Phillips $Z_t$ test statistics (p-value) | -6.46 (0.000) | -7.05 (0.000) | -6.93 (0.000) | -7.04 (0.000) | -6.59 (0.000) | -10.11 (0.000) |
| Estimated Break year | 2010 | 2011 | 2011 | 2011 | 2012 | 2013 |

### 5.7  Leave-one-out analysis

Table 6 reports leave-one-out analysis of the impact of the populist constitutional reforms on judicial independence outcomes. The general thrust of the leave-one-out analysis is the exclusion of the donor unit with the largest weight share in seeking to gauge the robustness of the estimated counterfactual scenario to the exclusion of the dominant donor. The failure to reproduce both the magnitude of the gap would indicate the presence of the outlying and disproportionate sway of the dominant donor and its potential outcome-related or auxiliary idiosyncrasies that may taint the counterfactual scenario with treatment-violating influences. The evidence from the leave-one-out analysis arguably suggests that piecewise exclusion of the dominant donor yields very similar impact magnitudes across the entire realm of the estimated specifications. In the court-packing model specification, the baseline result implies that Türkiye's synthetic control group is dominated by West Bank (33 percent of the total) whilst the end-of-sample gap is around 2 basis points. By discarding the West Bank from the donor pool, Türkiye's trajectory of court packing is best reproduced by a convex combination of the overall attributes of Mauritania (40%), Greece (31%), and Jordan (29), respectively. The end-of-sample court packing gap between Türkiye and its alternate synthetic control group is around -2.5 basis points and is remarkably consistent and almost identical to the baseline result. Similar evidence is found for the other judicial independence outcomes whereupon splitting the most influential country off the donor pool entails no difference with respect to the magnitude of the reforms' impact as well as to the composition of synthetic control groups. The alternative synthetic control groups also provide a very good quality of the fit and indicate no discrepancy in RMSE between the leave-one-out specification and the baseline quality of fit. Figure 9 presents leave-one-out estimated effect of constitutional reforms on judicial independence for each designated outcome.

**Figure 9**: Leave-one-out analysis of the effect of constitutional backsliding on judicial independence in Türkiye, 1987-2023

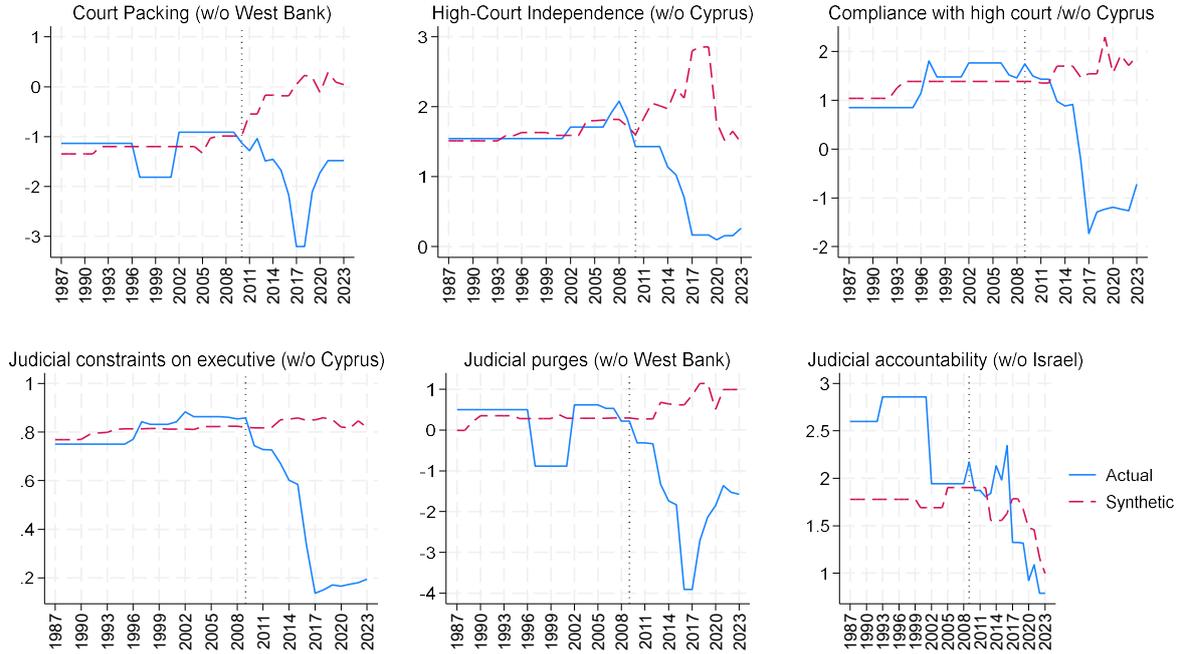

**Table 6**: Leave-one-out composition of synthetic control groups

|  | Court Packing | High-Court Independence | Compliance with High Court | Judicial Constraints on Executive | Judicial Purges | Judicial Accountability |
|---|---|---|---|---|---|---|
|  | w/o West Bank | w/o Cyprus | w/o Cyprus | w/o Cyprus | w/o West Bank | w/o Israel |
| RMSE | 0.330 | 0.094 | 0.301 | 0.039 | 0.605 | 0.982 |
| Albania | 0 | 0 | 0 | 0.11 | 0 | 0 |
| Algeria | 0 | 0 | 0 | 0 | 0 | 0 |
| Cyprus | 0 | excluded | excluded | excluded | 0 | 0.61 |
| Greece | 0.31 | 0 | 0 | 0 | 0 | 0 |
| Israel | 0 | 0 | 0 | 0.75 | 0 | excluded |
| Italy | 0 | 0 | 0 | 0 | 0 | 0 |
| Jordan | 0.29 | 0.38 | 0 | 0 | 0.76 | 0.20 |
| Malta | 0 | 0 | 0 | 0 | 0 | 0 |
| Mauritania | 0.40 | 0 | 0 | 0 | 0 | 0 |
| Morocco | 0 | 0 | 0 | 0 | 0.24 | 0 |
| North Macedonia | 0 | 0 | 0 | 0 | 0 | 0 |
| Portugal | 0 | 0.35 | 0.55 | 0 | 0 | 0 |
| Slovenia | 0 | 0 | 0 | 0 | 0 | 0 |
| Spain | 0 | 0 | 0 | 0 | 0 | 0.18 |
| West Bank | excluded | 0.28 | 0.45 | 0.14 | excluded | 0 |

### 5.8 Bias-corrected estimates

Lastly, Figure 10 reports the synthetic control estimates corrected for the biases arising from possibly non-unique solution to the cross-similarity matrix that reproduces the attributes of Türkiye prior to the constitutional reforms. By making use of Abadie and L'Hour (2021) penalized synthetic control estimator that penalizes these pairwise discrepancies, we correct our estimates by embedding a penalized parameter that trades off pairwise matching dissimilarities with respect to the implied attributes of each country in Türkiye's synthetic control group for each outcome considered in the vector of dependent variables. By penalizing the synthetic control estimates with the discrepancy-related penalization parameter, both sparsity and uniqueness conditions for the solution to find the optimal weights for Türkiye are satisfied which allows us to adjust the post-intervention judicial independence gaps through the correction of biases for the underlying treatment effects. In terms of further detail, bias-corrected estimates are obtained by fitting our model using both elastic net regularization that unifies and additionally improves the penalizes of the LASSO and ridge methods.

The evidence suggests a striking similarity between our baseline synthetic control estimated judicial independence effects of populist constitutional reforms. In particular, the classic synthetic control estimates and their bias-corrected counterparts are nearly identical in both pre-treatment and post-treatment period. Across the entire spectrum of outcome-level specifications, we find evidence of large-scale erosion of judicial independence although at different speed. For instance, the erosion of independence appears to be particularly faster and more pronounced through heighted politically-motivated judgeship appointment, retraction of judicial constraints on the executive and widespread onset of judicial purges. By contrast, the erosion of high-court independence and government's compliance with the high court appears to be somewhat more gradual and persistent whilst a drop in judicial accountability is seldom perceptible. The correlation between classic synthetic control estimates and their bias-corrected counterparts is in excess of +0.96 for the pooled outcome variables and is statistically significant at 1%, further reinforcing our baseline results.

**Figure 10**: Classical and bias-corrected effect of constitutional backsliding on judicial independence in Türkiye using synthetic control estimator, 1987-2023

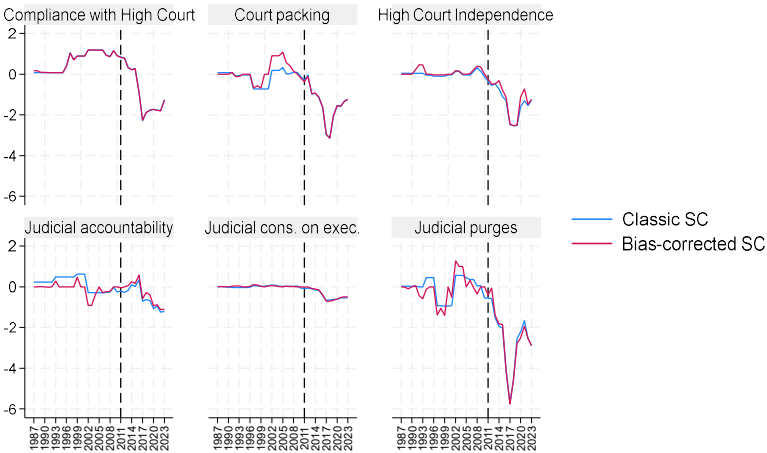

### 5.9 Effects under countercyclical weights

The estimated counterfactual scenario of constitutional backsliding is based on the selection of weights using the seminal approach of Abadie et. al. (2010, 2015) where several restrictions not required by identifying assumptions are imposed. One of the most salient restrictions requires that the synthetic control groups are composed using non-negative and additive weights which yields a convex combination of donor groups that best reproduce and track the trajectories of judicial independence of Türkiye prior to the constitutional reforms. In turn, the implicit attributes of Türkiye in the pre-intervention period emanate from the extrapolation inside its convex hull whilst preventing the extrapolation outside the convex hull. Recent advances in the literature on synthetic control method proposed a series of strategies for estimating counterfactual weights (Doudchenko and Imbens 2016, Powell 2019, Arkhangelsky et. al. 2021) based on the extrapolation outside the convex hull. To estimate counterfactual weights from the donor pool, we use LASSO-based extension of the synthetic control method (Hollingsworth and Wing 2020). Compared to the classical synthetic control estimator, LASSO extension is flexible and relies on the machine-learning approach to construct weights both inside and outside the sample allowing for countercyclical weights. By relaxing the convexity restriction, the improvement in the quality of the fit between Türkiye and its synthetic peer becomes straightforward.

Figure 11 presents the estimated effect of 2011 constitutional reform on judicial independence and the related outcomes. It becomes apparent that allowing for the extrapolation outside the convex hull of Türkiye's judicial attributes substantially improves the quality of the fit. For instance, in our preferred specification of high-court independence dynamics, LASSO-based synthetic control estimator exhibits 23 percent lower RMSE alongside a similar magnitude of improvement in the specification of compliance dynamics. A notable improvement is perceptible in judicial accountability dynamics where relaxing the convexity restrictions yields 41 percent lower RMSE compared to the classical synthetic control estimator. The evidence confirms our prior results. In particular, the trajectories of judicial independence, compliance and constraints tend to deteriorate substantially in the post-referendum period. The erosion of judicial independence and integrity is further pinpointed by rapid escalation of court-packing. LASSO-based synthetic control estimates also indicate a more gradual but permanent erosion of judicial accountability vis-á-vis the synthetic control group.

**Figure 11**: LASSO-estimated effects of constitutional overhaul of judicial independence in Türkiye, 1987-2023

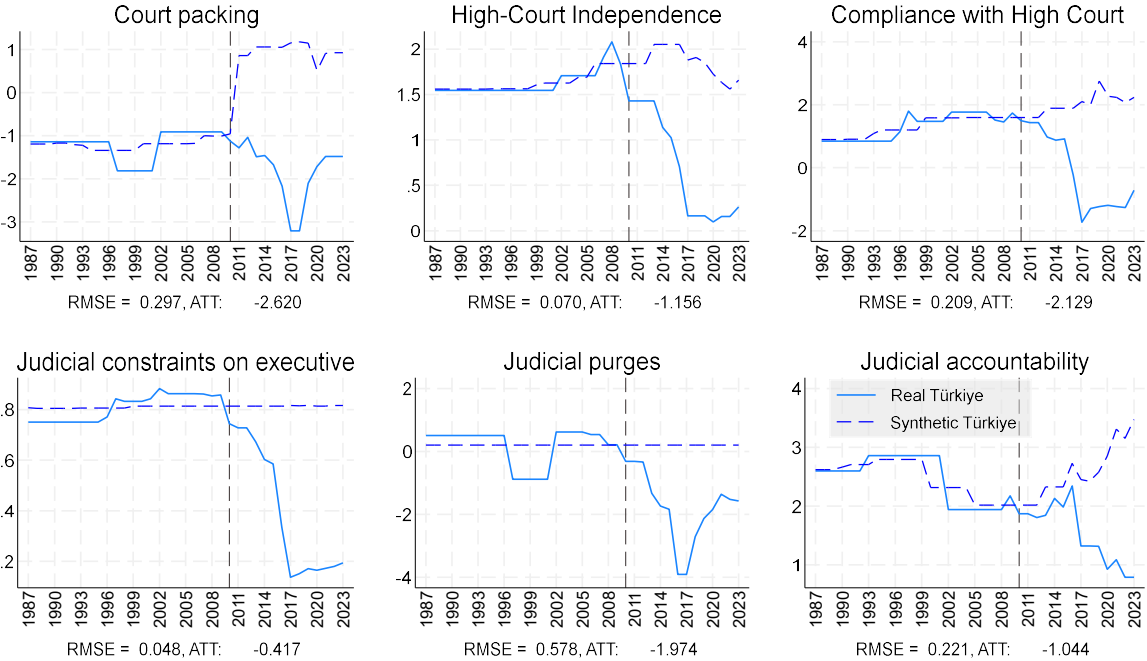

Table 7 reports the magnitude of the average treatment effect of the 2011 constitutional overhaul alongside the standard errors, two-tailed 95% empirical confidence intervals and post-intervention RMSE. Furthermore, Panel B reports non-convex composition on the synthetic control groups in greater detail. For instance, the synthetic version of Türkiye's court-packing trajectory prior to 2011 loads positively on Jordan (+1.17) and Cyprus (+0.09) whilst negatively on Israel (-0.03). The structure and composition of weights does not appear to be created uniformly across the full set of outcomes. For instance, in the high-court independence simulation, the synthetic version of Türkiye consists of the implied attributes of Cyprus (+0.23) and Mauritania (+0.22) whilst the only entity exhibiting a minor negative weight is the West Bank (-0.001). It should be noted that the application of the LASSO-based synthetic control estimator improves the quality of the fit for those outcomes in which convexity requirement yields null effect of constitutional referendum and subsequent overhaul such as judicial accountability. Allowing for countercyclical weights extrapolated outside the convex hull, LASSO synthetic control estimates indicate a gradual and persistent retrogression of judicial accountability up to the present day. Prior to the constitutional referendum, Türkiye's trajectory of judicial accountability is best reproduced as a non-convex combination of the implied attributes of three Mediterranean donor states, namely, Jordan (+1.104), Slovenia (+0.168) and Cyprus (-0.85).

Table 7: Treatment effect of constitutional referendum on judicial independence and integrity without convexity requirement in Türkiye, 1987-2022

|  | Court Packing | High-Court Independence | Compliance with High Court | Judicial Constraints on Executive | Judicial Purges | Judicial Accountability |
|---|---|---|---|---|---|---|
| Panel A: Average treatment effect | | | | | | |
| ATT | -2.620 | -1.156 | -2.129 | -.417 | -1.974 | -1.044 |
|  | (.296) | (.069) | (.208) | (.048) | (.578) | (.221) |
| 95% Confidence Interval | (-4.095, -1.146) | (-1.794, -.519) | (-3.328, -.929) | (-.649, -.184) | (-3.07, -.879) | (-1.628, -.46) |
| p-value | 0.000 | 0.000 | 0.000 | 0.000 | 0.000 | 0.000 |
| Post-Intervention RMSE | 2.804 | 1.258 | 2.573 | 0.485 | 2.260 | 1.366 |
| Panel B: Composition of synthetic control groups | | | | | | |
| RMSE | 0.330 | 0.094 | 0.301 | 0.039 | 0.605 | 0.221 |
| Albania | 0 | 0 | 0 | 0 | 0 | 0 |
| Algeria | 0 | 0 | -0.421 | -0.066 | 0 | 0 |
| Cyprus | 0.091 | 0.229 | 0.036 | 0 | 0 | -0.850 |
| Greece | 0 | 0 | 0 | 0 | 0 | 0 |
| Israel | -0.033 | 0 | 0 | 0 | 0 | 0 |
| Italy | 0 | 0 | 0 | 0 | 0.20 | 0 |
| Jordan | 1.175 | 0 | 0 | 0 | 0 | 1.104 |
| Malta | 0 | 0 | 0 | 0 | 0 | 0 |
| Mauritania | 0 | 0.227 | 0 | 0 | 0.04 | 0 |
| Morocco | 0 | 0 | 0 | 0 | 0 | 0 |
| North Macedonia | 0 | 0 | 0 | 0 | 0 | 0 |
| Portugal | 0 | 0 | 0 | 0 | 0 | 0 |
| Slovenia | 0 | 0 | 0 | 0 | 0 | 0.168 |
| Spain | 0 | 0 | 0 | 0 | 0 | 0 |
| West Bank | 0.104 | <0.01 | 0.381 | 0 | 0.76 | 0 |

### 5.10 Effects using lagged outcomes

To capture the similarity between Türkiye and its Mediterranean donor states, we use the level of judicial variables in benchmark year together with the auxiliary covariates to ensure a stable variance in the latent factor model that reproduce the respective trajectories of Türkiye prior to the constitutional overhaul inasmuch as possible. Recent methodological work in the synthetic control method has avidly demonstrated that the use of lagged outcome values may further improve both the performance of synthetic control estimator as well as the quality of the fit in the pre-intervention period. Although lagged levels of the outcome variables may render the diagonal weight of the auxiliary covariates close to zero (Doudchenko and Imbens 2016), the reduction of imbalance between the trajectories of Türkiye and its synthetic peer can non-trivially improve the inference on the treatment effect.

To address the potential caveat in the empirical analysis, Figure 12 reports the effects of constitutional overhaul on judicial outcomes using two lags of each outcome variable instead of the values in the benchmark year. The evidence confirms our prior results and provides further empirical support to the notion that the 2011 constitutional overhaul dampened judicial independence through the erosion of constraints, annihilation of government's compliance with the Supreme Court and widespread escalation of both politically-motivated appointments (i.e. court packing) and judicial purges. The correlation between our baseline estimates and the estimates using two lags of the outcome variables is +0.99 and statistically significant at 1% (i.e. p-value = 0.000). In the lieu of the very similar composition of the outcome-specific synthetic control groups, it should also be noted that the reliance on lagged outcomes tends to diminish the outcome comparison bias between Türkiye and its synthetic peer below 1 percent, respectively. Lastly, Figure 13 also presents the effects of constitutional overhaul using the synthetic matching between Türkiye and its Mediterranean donor states based on the full path of pre-treatment outcome as suggested in the recent literature (Kaul et. al. 2022, Keseljevic and Spruk 2024). The estimated gaps in judicial outcome variables are consistent with our baseline estimates and prior robustness checks. In particular, the correlation between the baseline estimates and full pre-$T_0$ outcome path estimates is +0.98 and statistically significant at 1% (i.e. p-value = 0.000).

**Figure 12**: Effects of constitutional overhaul on judicial independence using dynamic latent factor model in Türkiye, 1987-2023

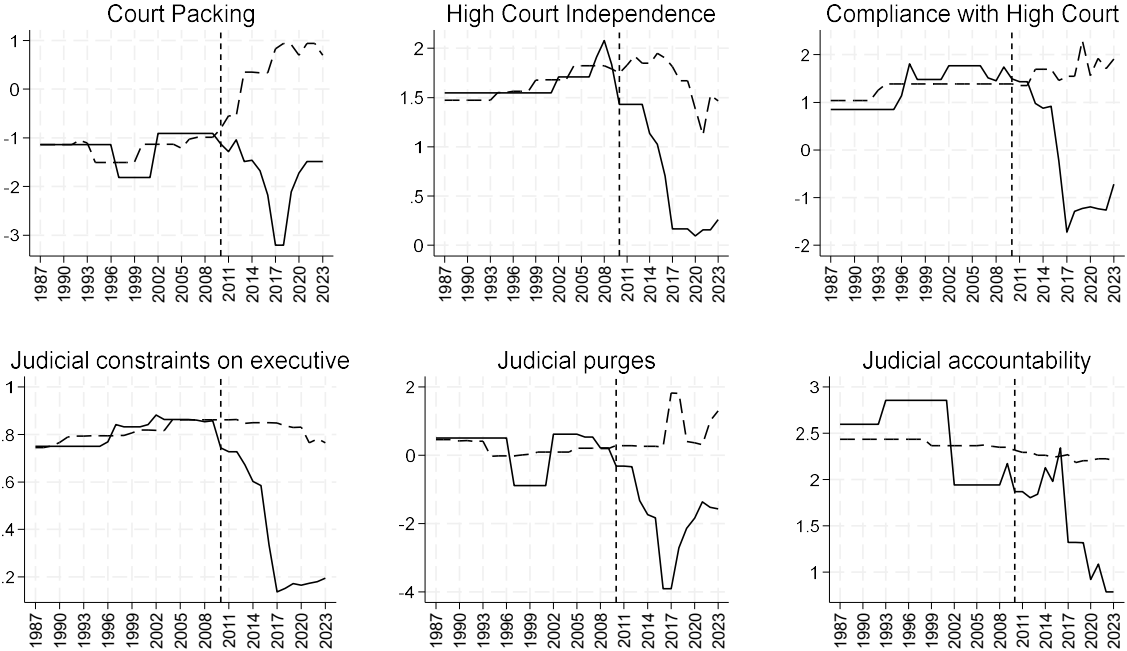

Table 8: Treatment effect of constitutional overhaul and the composition of synthetic control groups using dynamic latent factor model

| | Court Packing | High-Court Independence | Compliance with High Court | Judicial Constraints on Executive | Judicial Purges | Judicial Accountability |
|---|---|---|---|---|---|---|
| | (1) | (2) | (3) | (4) | (5) | (6) |
| Panel A: Treatment effect and bias | | | | | | |
| ATT | -2.313 | -1.049 | -1.932 | -0.459 | -2.553 | -1.445 |
| (RMSPE) | (.288) | (.094) | (.301) | (.031) | (.525) | (.374) |
| Lagged outcome bias (%) | <1% | <1% | <1% | <1% | <1% | <1% |
| p-value | 0.017 | 0.154 | 0.036 | 0.052 | 0.011 | 0.083 |
| Panel B: Convex composition of synthetic control groups | | | | | | |
| Albania | 0 | 0 | 0 | 0.10 | 0.47 | 0 |
| Algeria | 0 | 0 | 0 | 0 | 0 | 0 |
| Cyprus | 0.227 | 0.59 | 0 | 0.873 | 0 | 0 |
| Greece | 0 | 0 | 0 | 0 | 0 | 0 |
| Israel | 0 | 0 | 0 | 0 | 0 | 0.937 |
| Italy | 0 | 0 | 0 | 0 | 0.156 | 0 |
| Jordan | 0.171 | 0 | 0 | 0 | 0 | 0 |
| Malta | 0 | 0 | 0 | 0 | 0 | 0 |
| Mauritania | 0.248 | 0 | 0 | 0 | 0 | 0 |
| Morocco | 0 | 0.145 | 0 | 0.027 | 0 | 0 |
| North Macedonia | 0 | 0 | 0 | 0 | 0 | 0.063 |
| Portugal | 0 | 0.265 | 0.55 | 0 | 0 | 0 |
| Slovenia | 0 | 0 | 0 | 0 | 0 | 0 |
| Spain | 0 | 0 | 0 | 0 | 0 | 0 |
| West Bank | 0.354 | 0 | 0.45 | 0 | 0.796 | 0 |

Figure 13: Effects of constitutional overhaul on judicial independence using full pre-$T_0$ outcome path without covariates in Türkiye, 1987-2023

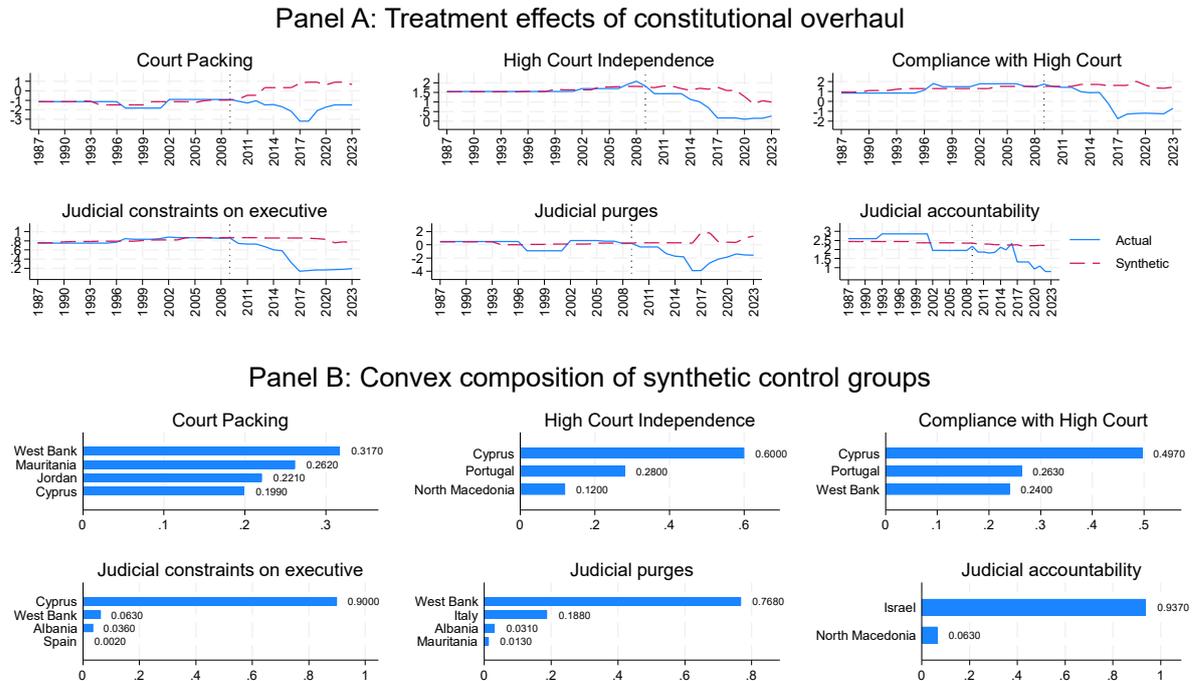

Panel A: Treatment effects of constitutional overhaul

Panel B: Convex composition of synthetic control groups

## 6   Conclusion

To bridge the gap between the theoretical literature and the emerging empirical applications of the synthetic control method, we examine the populist constitutional reforms implemented by Turkish President Recep Tayyip Erdoğan following the 2010 constitutional referendum. These reforms have widely been regarded as a critical turning point, marking the onset of populist regression and the deterioration of judicial independence. Using the recently updated Varieties of Democracy dataset and a large sample of countries from 1987 to 2023, we estimate the counterfactual scenario of Türkiye's judicial independence in the hypothetical absence of the populist backsliding that began in 2010.

Our findings show that the 2010 referendum triggered a sustained and irreversible erosion of judicial independence, characterized by heightened politically motivated judicial appointments, widespread purges, a significant weakening of judicial constraints on government, and the government's increasing disregard for judicial rulings. This decline in judicial independence is robust to a range of placebo tests and other robustness checks, and it persists through the present day. Moreover, generating a counterfactual using pre-referendum variation in interactive fixed effects further supports our conclusions. We find little evidence to suggest that the 2017 constitutional amendments played a significant role in the subsequent erosion of judicial independence, reinforcing the idea that the pivotal break occurred in 2010.

The synthetic control method offers numerous opportunities for its application in empirical comparative law, enabling scholars and practitioners to estimate missing counterfactual scenarios in response to significant changes in legal systems, constitutions, election laws, policies, or other institutional reforms. When a well-constructed research design—featuring clearly defined treatment and control groups—is paired with the battery of tests and strategies outlined in this study, scholars can generate policy-relevant and insightful evidence on the effects of constitutional and legal changes. This method not only helps elucidate the causal impacts of such interventions but also sheds light on their broader policy implications.